\documentclass[manuscript, screen, nonacm, natbib=false]{acmart}
\usepackage[utf8]{inputenc}
\usepackage{enumitem}
\usepackage[most]{tcolorbox}
\usepackage{hyperref}
\usepackage{xcolor}
\definecolor{grey}{rgb}{0.5,0.5,0.5}
\usepackage{tcolorbox}
\usepackage{array}
\usepackage{listings}
\usepackage{tabularx}
\usepackage{multirow}
\usepackage{graphicx}
\usepackage{subcaption}
\usepackage{soul}
\usepackage{amsmath}
\usepackage{tikz}
\usepackage[utf8]{inputenc}
\DeclareUnicodeCharacter{064E}{}

\AtBeginDocument{%
  }

\newcommand \tool{LegoUI}

\RequirePackage[
  datamodel=acmdatamodel,
  style=acmnumeric,
  ]{biblatex}

\newcolumntype{Y}{>{\raggedright\arraybackslash}X} 
\renewcommand{\arraystretch}{1.1}                   

\newtcbox{\blue}{on line, boxrule=0pt, colback=blue!15, colframe=blue!40,
  boxsep=1pt, left=2pt, right=2pt, top=1pt, bottom=1pt, arc=3pt, box align=base,
  coltext=blue!80!black, fontupper=\ttfamily\bfseries}

\newtcbox{\green}{on line, boxrule=0pt, colback=green!15, colframe=green!40,
  boxsep=1pt, left=2pt, right=2pt, top=1pt, bottom=1pt, arc=3pt, box align=base,
  coltext=green!80!black, fontupper=\ttfamily\bfseries}

\newtcbox{\orange}{on line, boxrule=0pt, colback=orange!15, colframe=orange!40, boxsep=1pt, left=2pt, right=2pt, top=1pt, bottom=1pt, arc=3pt, box align=base, coltext=orange!80!black, fontupper=\ttfamily\bfseries, breakable, tcbox raise base}

  \newtcbox{\teal}{on line, boxrule=0pt, colback=teal!15, colframe=teal!40,
  boxsep=1pt, left=2pt, right=2pt, top=1pt, bottom=1pt, arc=3pt, box align=base,
  coltext=teal!80!black, fontupper=\ttfamily\bfseries}

\newtcbox{\grey}{on line, enhanced jigsaw, breakable,
  boxrule=0pt,
  colback=gray!15,     
  colframe=gray!40,    
  boxsep=1pt,
  left=2pt, right=2pt, top=1pt, bottom=1pt,
  arc=3pt,
  box align=base,
  coltext=black,       
  fontupper=\ttfamily\bfseries  
}

\newtcbox{\lavender}{on line, boxrule=1pt, 
  colback=white,          
  colframe=violet!40,     
  boxsep=1pt, left=2pt, right=2pt, top=1pt, bottom=1pt, 
  arc=3pt, box align=base,
  coltext=violet!80!black, 
  fontupper=\ttfamily\bfseries}

\definecolor{softblue}{HTML}{6385AC}
\newtcbox{\lightblue}{on line, boxrule=1pt, 
  colback=white,          
  colframe=softblue!80,   
  boxsep=1pt, left=2pt, right=2pt, top=1pt, bottom=1pt, 
  arc=3pt, box align=base,
  coltext=softblue!80!black,  
  fontupper=\ttfamily\bfseries}

\newtcbox{\purple}{on line, boxrule=0pt, 
  colback=white,        
  colframe=white,       
  boxsep=0pt, left=0pt, right=0pt, top=0pt, bottom=0pt, 
  arc=0pt, box align=base,
  coltext=purple!80!black, 
  fontupper=\ttfamily\bfseries}

\newtcblisting{CodeBlock}{
  enhanced,
  colback=grey!15, colframe=gray!40,
  boxrule=0pt, arc=3pt, outer arc=3pt,
  boxsep=4pt, left=3pt, right=3pt, top=3pt, bottom=3pt,
  breakable,
  listing only,
  listing engine=listings,
  listing options={
    basicstyle=\ttfamily\small,
    breaklines=true,
    columns=fullflexible,   
    keepspaces=true,        
    showstringspaces=false  
    frame=none,                           
     backgroundcolor=\color{grey!15} 
  }
}

\definecolor{codegray}{RGB}{245,245,245}
\definecolor{deepblue}{HTML}{261F4B}

\usepackage{tcolorbox}
\tcbuselibrary{listings,skins,breakable}
\tcbset{
  colback=gray!5,
  colframe=black!20,
  boxrule=0.3pt,
  left=6pt,right=6pt,top=6pt,bottom=6pt,
  sharp corners,
  breakable
}

\newcommand{\circled}[2]{%
  \tikz[baseline=(char.base)]{
    \node[
      circle,
      fill=#1,
      draw=#1,
      inner sep=1pt,
      minimum size=1.0em
    ] (char) {\textcolor{white}{\small\bfseries #2}};
  }%
}

\renewcommand\footnotetextcopyrightpermission[1]{}

\author{Yinsi Zhou}
\orcid{0009-0009-5675-3937} 
\affiliation{%
  \institution{School of Computer Science and Engineering, University of New South Wales}
  \city{Sydney}
  \state{NSW}
  \country{Australia}
}

\author{Mingyue Yuan}
\orcid{0009-0004-5797-0945}
\authornote{Corresponding author: Mingyue Yuan, mingyue.yuan@unsw.edu.au.}
\affiliation{%
  \institution{School of Computer Science and Engineering, University of New South Wales}
  \city{Sydney}
  \state{NSW}
  \country{Australia}
}

\affiliation{%
  \institution{Data61, CSIRO}
  \city{Eveleigh}
  \state{NSW}
  \country{Australia}
}

\author{Hongyue Xu}
\orcid{0009-0000-9771-7961}
\affiliation{%
  \institution{Interactive Content Design Lab, Tohoku University}
  \city{Sendai}
  \state{Miyagi}
  \country{Japan}
}

\author{Jieshan Chen}
\orcid{0000-0002-2700-7478}
\affiliation{%
  \institution{Data61, CSIRO}
  \city{Eveleigh}
  \state{NSW}
  \country{Australia}
}

\author{Dong Wen}
\orcid{0000-0002-0903-1503}
\affiliation{%
  \institution{School of Computer Science and Engineering, University of New South Wales}
  \city{Sydney}
  \state{NSW}
  \country{Australia}
}

\author{Shidong Pan}
\orcid{0000-0002-2162-0407}
\affiliation{%
  \institution{Center for Data Science, New York University}
  \city{New York}
  \state{NY}
  \country{United States}
}
\email{shidong.pan@nyu.edu}

\author{Xiwei Xu}
\orcid{0000-0002-2273-1862}
\affiliation{%
  \institution{Data61, CSIRO}
  \city{Eveleigh}
  \state{NSW}
  \country{Australia}
}

\author{Wenjie Zhang}
\orcid{0000-0001-6572-2600}
\affiliation{%
  \institution{School of Computer Science and Engineering, University of New South Wales}
  \city{Sydney}
  \state{NSW}
  \country{Australia}
}

\author{Aaron Quigley}
\orcid{0000-0002-5274-6889}
\affiliation{%
  \institution{College of Systems and Society, Australian National University}
  \city{Canberra}
  \state{ACT}
  \country{Australia}
}

\author{Zhenchang Xing}
\orcid{0000-0001-7663-1421}
\affiliation{%
  \institution{Data61, CSIRO}
  \city{Eveleigh}
  \state{NSW}
  \country{Australia}
}
\affiliation{%
  \institution{College of Engineering, Computing and Cybernetics, Australian National University}
  \city{Canberra}
  \state{ACT}
  \country{Australia}
}

\author{Gelareh Mohammadi}
\orcid{0000-0002-8087-2241}
\affiliation{%
  \institution{School of Computer Science and Engineering, University of New South Wales}
  \city{Sydney}
  \state{NSW}
  \country{Australia}
}

\makeatletter
\def\@authorsaddresses{}
\makeatother

\begin{document}

\title{LEGOUI: Designing with UI-DSL Bricks to Balance Transparency and Controllability}

\renewcommand{\shortauthors}{Zhou et al.}


\begin{abstract}
Generative user interface design tools enable rapid prototyping but often operate as black boxes with limited transparency and controllability. When outputs diverge from the designer’s intent, users are left tweaking prompts via trial-and-error with little insight into the model’s reasoning. We present \tool{}, a staged generative framework that structures the interface design process into sequential, interpretable steps along key design dimensions, capturing each step’s result in a UI domain-specific language (UI-DSL) enriched with provenance. This approach exposes the model’s intermediate reasoning and enables user intervention and iterative refinement. In a technical evaluation on 40 real-world design prompts, \tool{}’s requirement analysis stage captured explicit requirements with over 95\% accuracy, near-complete coverage, and zero redundancy. In user studies, participants using \tool{} reported significantly greater transparency, controllability, and alignment with their intent compared to existing one-shot generative UI tools.
\end{abstract}



\maketitle

\begin{figure}[t]
  \centering
  \includegraphics[width=\linewidth]{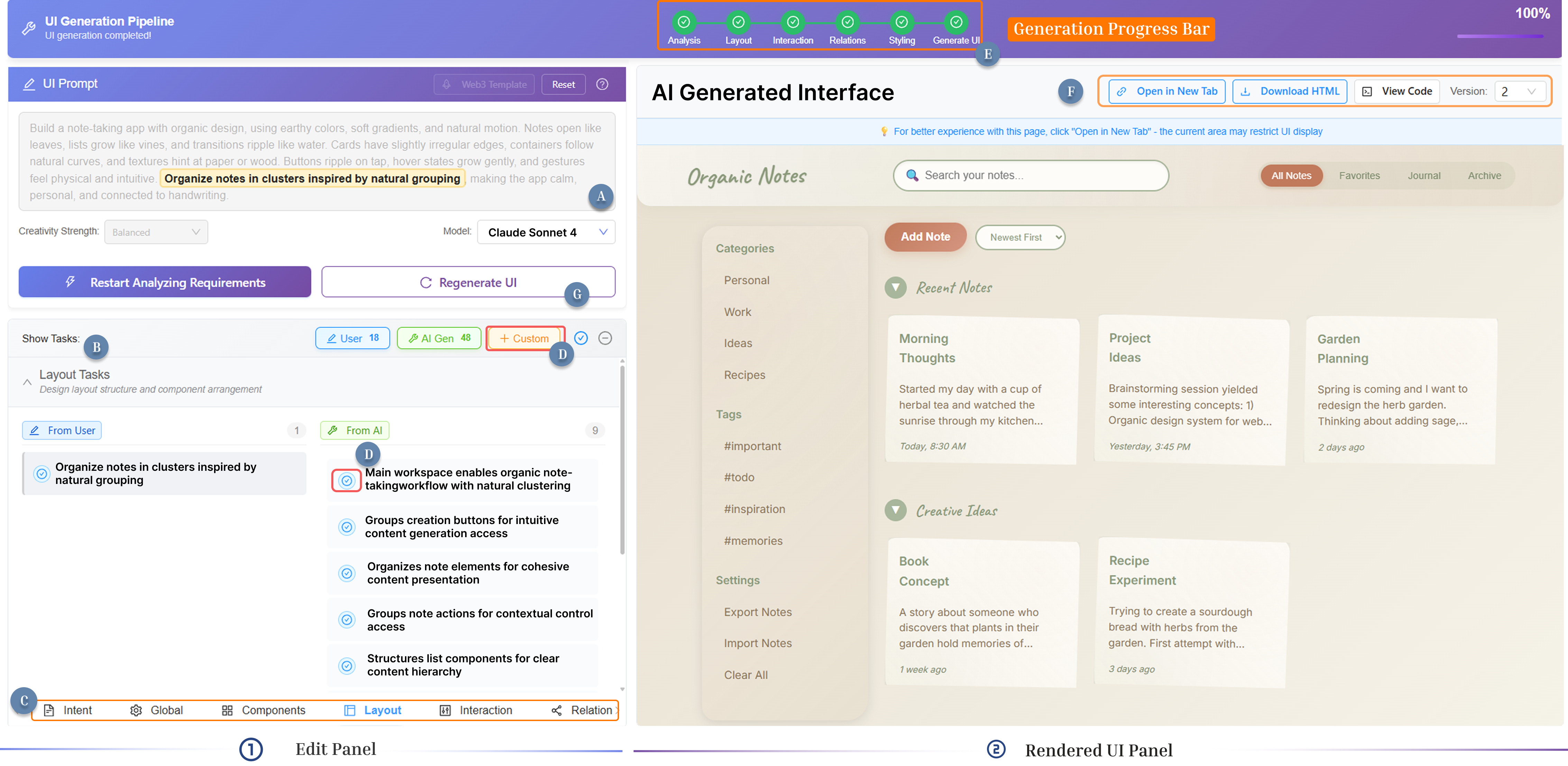}
  \caption{\textbf{The LegoUI system for staged UI generation.}
  LegoUI starts from a user prompt \protect\circled{softblue}{A} and incrementally constructs a UI specification through successive reasoning stages. Inferred design decisions are written into a persistent UI-DSL and surfaced as editable tasks \protect\circled{softblue}{B}, organized by design dimensions \protect\circled{softblue}{C}. Users intervene on these decisions through explicit actions \protect\circled{softblue}{D}, which update the active specification state that guides subsequent inference. The generation process advances through staged reasoning phases \protect\circled{softblue}{E}, and the current specification can be rendered at any point to produce an interface artifact \protect\circled{softblue}{F}. Further generation continues from the existing specification state \protect\circled{softblue}{G}, allowing interface variants to reflect accumulated design intent over time.}
  \label{fig:teaserfigure}
\end{figure}

\section{Introduction}

Generative user interface (GenUI) systems increasingly enable rapid production of high-fidelity interfaces from natural language descriptions and design artifacts.
Recent work shows that large language models can synthesize executable interface structures with limited upfront effort, supporting early-stage exploration across diverse application contexts
\cite{anthropic_artifacts_2024,bolt_new_2025,lovable_dev_2025,v0_app_2025,gui2025uicopilot,si2024design2code,kolthoff2024zero}.

These capabilities are especially attractive for beginner users during early design ideation.
Beginner users may have a sense of the interface they want to create, yet have limited experience translating that intent into concrete design decisions such as components, layout structure, interaction behavior, and visual hierarchy.
A prompt-to-UI system can help them begin quickly, but it can also make design decisions on their behalf without showing how those decisions were formed.
When the generated interface diverges from their intent, beginner users may struggle to identify whether the mismatch came from their prompt, the model's interpretation, or the design assumptions introduced during generation.

Most existing GenUI systems formulate interface generation as an end-to-end transformation from user intent to a complete interface artifact.
Under this formulation, generation is treated as a single operation, and the intermediate design decisions that guide the result are neither explicitly represented nor persistently maintained.
As a consequence, users primarily interact with such systems through prompt-based trial and error, repeatedly regenerating full interfaces to approximate desired outcomes~\cite{dang2022prompt,petridis2023promptinfuser}.
For beginner users, this makes iteration particularly difficult because prompt revision becomes an indirect way to debug design assumptions that are hidden inside the output.
This interaction paradigm is further constrained by a mismatch between unstructured natural language and the structured constraints required in UI design, such as component variants, fine-grained styling, and cross-component consistency
\cite{chen2025genui,zamfirescu2023johnny,lu2025misty,shokrizadeh2025dancing}.

UI design practice unfolds through successive refinement as requirements evolve from abstract goals to concrete decisions, particularly during early ideation.
Beginner users often form these judgments while seeing the design develop.
They may need to confirm some assumptions, revise others, and gradually decide which aspects of the interface should remain stable.
In current GenUI systems, iteration typically manifests as repeated full-interface generation, which repeatedly reinterprets prior intent, increases variability across outputs, and makes it difficult to preserve and build upon earlier design decisions
\cite{lee2024one,elazar2021measuring,v0_app_2025,bolt_new_2025}.
These limitations become particularly pronounced in longer-term and iterative design settings, where generated interfaces are expected to conform to shared visual and interaction standards and integrate with existing design systems.

Recent systems introduce interaction mechanisms that allow users to influence generation during execution
\cite{cao2025generative,beason2025athena,leung2025squire}.
These approaches demonstrate the value of exposing intermediate structure and creating more interactive forms of GenUI.
At the same time, the resulting decisions are often transient or tied to a specific generation episode.
Even when interaction is supported, existing systems provide limited support for persisting, revising, and accumulating intermediate design decisions as requirements evolve over time
\cite{chen2025genui}. 
Table~\ref{tab:genui-comparison} situates our approach relative to these strands of GenUI research.

\begin{table}[t]
\centering
\small
\setlength{\tabcolsep}{5pt}
\renewcommand{\arraystretch}{1.25}
\begin{tabular}{p{0.20\columnwidth}p{0.17\columnwidth}p{0.17\columnwidth}p{0.17\columnwidth}p{0.17\columnwidth}}
\toprule
\textbf{Approach} &
\textbf{Explicit intermediate decisions} &
\textbf{Selective control over decisions} &
\textbf{Evolving intent across iterations} &
\textbf{Long-term reuse across pages/time} \\
\midrule

End-to-end prompt-based GenUI
\cite{anthropic_artifacts_2024,v0_app_2025,bolt_new_2025,lovable_dev_2025}
&
No explicit representation
&
Prompt-only control
&
Reinterpreted each generation
&
No reuse support
\\

Interactive GenUI with intermediates
\cite{beason2025athena,cao2025generative,chen2025specifyui,leung2025squire}
&
Session-scoped representations
&
Scoped structure edits
&
Session-level continuity
&
Session-scoped reuse
\\

\addlinespace[4pt]
\textbf{Our approach}
&
Persistent staged specification
&
Decision-level edits across iterations
&
Incremental intent accumulation
&
Cross-page reuse support
\\

\bottomrule
\end{tabular}
\caption{Comparison of representative GenUI approaches in terms of mechanisms for intermediate decisions, control, iteration, and reuse.}
\label{tab:genui-comparison}
\end{table}

Motivated by these challenges, we introduce a staged generation framework that supports UI synthesis as an iterative co-design process for beginner users, particularly during early design ideation where requirements are gradually clarified through interaction with the system.
The framework externalizes intermediate design decisions into a structured specification that can be incrementally refined across stages.
This allows beginner users to inspect how the system interprets their prompt, decide which inferred assumptions should remain active, and preserve stable design commitments as other decisions evolve.
To support this process, we propose a UI domain-specific language (UI-DSL) that represents interface structure, design constraints, and provenance as editable specification items.
The UI-DSL makes AI-inferred design decisions available for user judgment before they are rendered into a final interface.

We instantiate this framework in \tool{}, a prototype system that integrates large language model reasoning with staged specification construction (Figure~\ref{fig:teaserfigure}).
Given a user prompt, \tool{} derives an initial UI-DSL specification and incrementally extends it across stages.
Users can accept, reject, or add specification items during this process, allowing their feedback to shape the reasoning state used for later generation.
Interface realizations are generated from the current specification state, enabling alternative designs to be explored without discarding accumulated intent.
This staged organization supports a shift from post-hoc correction of generated artifacts toward earlier intervention in the assumptions that guide generation.

We evaluate this approach through a technical evaluation and a user study with 15 participants.
The technical evaluation examines the ability of large language models to derive structured specifications from natural language prompts.
The user study compares \tool{} with existing GenUI systems and investigates how staged generation shapes beginner users' perceived control, understanding, output alignment, and iteration strategies during interface design.
Our findings show that staged specification can make prompt interpretation easier to inspect, support earlier and more targeted intervention, and improve perceived alignment and coherence.
They also show that inspectable generation introduces additional interaction cost, especially when users need to evaluate abstract or dependency-based decisions.

This paper makes the following contributions:
\begin{itemize}
    \item A staged GenUI framework that treats prompt-to-interface generation as an iterative process of inspecting and revising AI-inferred design commitments for beginner users.
    \item A UI-specific domain-specific language that externalizes intermediate design decisions as provenance-aware items that can be inspected, accepted, rejected, and revised across stages.
    \item A prototype system, \tool{}, and empirical findings showing how staged intermediate representations affect beginner users' perceived transparency, control, output alignment, and interaction cost compared with one-shot GenUI systems.
\end{itemize}
\section{Related Work}

\subsection{GUI Design Tools and Generative Techniques}

Early work in interface generation emphasised specification-driven pipelines, where declarative rules could be compiled into executable prototypes~\cite{schreiber1994specification}. This paradigm extended to visualization grammars such as Vega-Lite~\cite{vega} and NL4DV~\cite{NL4DV}, which lowered barriers for analytic interfaces through high-level specifications. In parallel, retrieval-based tools such as Guigle~\cite{Guigle}, Swire`\cite{swire}, VINS~\cite{VINS}, Gallery D.C.~\cite{Gallery}, and Wireframe Autoencoder~\cite{Wireframe} enabled designers to search large corpora using text, sketches, or structural similarity, providing inspiration but offering limited support for novel layouts.
With advances in machine learning, representation learning approaches have sought to capture latent structures of interfaces\cite{li2021screen2vec,li2020widget}, offering reusable semantic building blocks that improve retrieval and accessibility. More recently, generative techniques have demonstrated how algorithmic control and adversarial modeling can synthesize layouts and components while supporting compositional creativity~\cite{cheng2023play,zhao2021guigan}. Systems such as PrototypeFlow~\cite{yuan2024towards} and Jelly~\cite{cao2025generative} introduced intermediate models to improve transparency and enable iterative modification, though limitations remain in reliability and mapping quality.

Parallel work has focused on code generation from visual or multimodal inputs. Early systems like pix2code~\cite{beltramelli2018pix2code} pioneered screenshot-to-code pipelines, while recent efforts such as UICoder~\cite{wu2024uicoder}, UICopilot~\cite{gui2025uicopilot}, DCGen~\cite{wan2024automatically}, and Web2Code~\cite{yun2024web2code} target modern webpage generation strategies. These studies reveal both the progress and the persistent challenges of multimodal LLMs in capturing layout fidelity and producing correct, editable structures.

Research has progressed from formal specifications and retrieval-driven inspiration to increasingly powerful generative methods. Yet existing approaches rarely distinguish the different concerns that arise across a design process. Our work addresses this gap by introducing stage-level reasoning into GUI generation. We structure generation into successive stages that mirror how designers move from intent to layout and interaction. This perspective makes the generative process more aligned with established practices of interface design.

\subsection{Human–AI Collaboration in Design Workflows}
The rise of generative AI has prompted researchers and practitioners to explore how these systems integrate into design workflows. Studies report both excitement about their creative potential and concerns over compatibility with established practices~\cite{lu2022bridging,chen2025genui,li2024user}. Current tools are valued for producing quick visual artifacts, yet fall short in higher-level activities such as user research or testing~\cite{lu2022bridging}. Designers often regard AI as an assistive partner, but also raise concerns about skill degradation and creativity fatigue~\cite{chen2025genui,li2024user}. Industry tools such as Lovable, Vercel v0, Bolt, and Claude can rapidly generate mockups or code, but their black-box nature makes them better suited as starting points than sustained collaborators. How to extend generative systems beyond one-off creation remains an open challenge.

A complementary line of work has focused on demonstration and hybrid interaction as mechanisms for human–AI collaboration. Rather than relying on one-shot generation, systems that learn from user demonstrations or combine multiple modalities allow co-construction of interfaces~\cite{vaithilingam2019bespoke,li2017sugilite}. 
Cao et al.~\cite{cao2025generative} highlight how direct manipulation of AI-generated artifacts can ground collaboration. Other systems extend demonstration with multimodal interaction. SOVITE~\cite{li2020multi} grounds conversational repair in GUI screenshots and direct manipulation, helping users resolve misunderstandings in task-oriented dialogue. DynaVis~\cite{vaithilingam2024dynavis} combines natural language input with dynamically generated widgets, allowing users to edit visualizations iteratively while maintaining feedback. Misty~\cite{lu2025misty} encourages developers to blend elements from sketches and screenshots into new prototypes, offering serendipitous combinations that promote creative exploration. These approaches suggest that AI’s role is best understood as that of a collaborative partner that provides starting points and adjustable artifacts, while humans iteratively adapt results to align with their goals.

Overall, prior work shows that effective human–AI collaboration in design requires iterative workflows, transparency, and opportunities for human steering. Building on these insights, our work embeds generative processes within a scaffolded, staged generation workflow to ensure that designers retain agency over intermediate representations and their refinements.

\subsection{LLM-Powered Design Assistance and Customization}

LLMs have recently been applied to interface design, enabling lightweight customization and automated feedback. For example, Stylette~\cite{kim2022stylette} supports natural language–driven styling changes, and Duan et al.~\cite{duan2024generating} introduced a Figma plugin that leverages GPT-4 for heuristic evaluation of mockups. These tools lower the barrier for non-experts and provide immediate suggestions, though their focus is often limited to specific aspects such as styling or usability, offering only partial coverage of broader design needs.

Meanwhile, evaluation benchmarks such as Design2Code~\cite{si2024design2code} and Web2Code~\cite{yun2024web2code} have revealed the fragility of current multimodal prompting, including frequent omission of visual elements and distorted layouts. 
Rather than treating these failures as end points, more recent efforts have sought to redesign the workflow itself. 
Multimodal modular frameworks~\cite{yuan2024towards} emphasize intent clarification, while Bespoke~\cite{vaithilingam2019bespoke} and Guide~\cite{GUIDE} illustrate how iterative prompting and editable specifications can scaffold designer control. Likewise, malleable interface models~\cite{cao2025generative} show how direct manipulation of generated artifacts can extend beyond one-shot generation. 
These works point to a promising direction: from output-focused generation toward dependable, iterative assistance that improves designer control.

Our work addresses these gaps by introducing explicit intermediate UI-DSL representations. Unlike black-box prompting, this structured approach makes model outputs traceable and adjustable, enabling designers to iteratively guide results and better align them with design intent.
\section{Design Goals}\label{sec:3_DG}

The GenUI Study~\cite{chen2025genui} highlights key shortcomings of existing generative UI tools. Building on these insights, particularly GenUI Gaps in  \ul{\textit{``Problem Formulation with Context''}}, \ul{\textit{``Assimilating Users’ Intents''}}, as well as \ul{\textit{``Quality, Fidelity, \& Originality''}} and \ul{\textit{``Support for Editing \& Iteration''}}. We aim to develop a generative, transparent, and user-controllable UI system. Our \tool{} emphasizes structured and traceable expression mechanism between user intent description and system inference, leading to the following design goals (DGs):

\begin{description}[leftmargin=2.8em, style=nextline, labelindent=0pt, labelwidth=2.5em]

\item[DG1] \textbf{Structured Prompts and Explicit Reasoning.}
Current generative UI systems often treat a user's natural language input as a single, opaque instruction, directly producing a final output in a black-box manner. While this approach can be effective when the outcome aligns with user expectations, it becomes problematic when it does not. 
The core issue is that this black-box process conceals the reasoning process: users cannot easily discern which aspects of the result come from their explicit input and which are inferred by the model. This lack of transparency limits users’ understanding of how the system interprets their intent and reduces their ability to refine or control the outcome through prompt design. To address this, we propose to structure user prompts and AI inference so that key elements of intent ( e.g., such as task goals, component requirements, and interaction preferences) are explicitly presented in distinct aspect. Furthermore, by separating what is provided by the user from what is inferred by the system, we makes AI reasoning visible, enabling users to trace, question, and adjust the generative process more effectively.

\item[DG2] \textbf{Staged Generation Process for Controllability.}
To enhance user understanding and control of the interface generation process, we propose to replace one-shot generation with a structured, staged reasoning process. Existing generation systems employ a single-round input-output mechanism, mixing design dimensions like layout, interaction, and style. This make it difficult for users to intervene in the intermediary stages and track the system's reasoning path. We aim to break down the the generation process into logical stages and each stage focuses on a specific design aspect such as component extraction, layout construction, interaction setting, etc. We also document the generation results and decision-making source through a structured representation.

\item[DG3] \textbf{Enabling Users to Customize Inference.}
During generation, the system may infer interface structures or components that the users has not explicitly specified. While such inferences can enrich the design, they may also diverge from user needs or expectations. To address this, users should be able to explicitly reject unwanted inferences and quickly supplement or modify design inference as desired. 
Direct customization enables users shape the interface and guide the system to better reflect their design intent.

\end{description}
\section{LegoUI Usage Example} \label{sec:exxample}

Emma is a design student participating in a workshop where she is asked to propose an interface for a blog page and justify her design decisions. The brief leaves key aspects of the interface open to interpretation, requiring participants to articulate their own design direction. To explore possible directions efficiently, Emma turns to generative tools during early-stage ideation. After interacting with tools that directly produce complete interfaces, Emma tries \tool{} on the same design task, starting from the same description.

\noindent\textbf{Getting Started:}
Emma begins by entering a concrete design prompt into \tool{} that captures her intended blog experience: “\textit{Design a blog interface in the spirit of editorial magazines, where oversized headlines and crisp excerpts pair with cover images and subtle tags....}” \circled{deepblue}{1}. She then selects a creativity setting and a generation model to proceed with the task  \circled{deepblue}{2}.

\begin{figure}[h]
    \centering
    \includegraphics[width=1\linewidth]{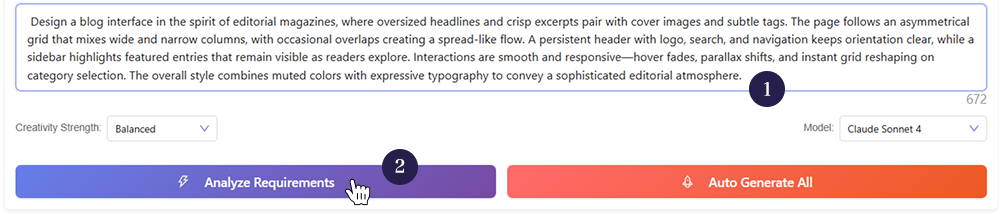}
    \label{fig:1-2}
\end{figure}

\noindent\textbf{Constructing the UI-DSL Specification: }
After submitting the prompt, LegoUI populates the Task Panel with a set of items derived from Emma’s description, each linked to a highlighted phrase in the prompt to show how her intent has been interpreted  \circled{deepblue}{3}.

Emma switches between tabs in the Task Panel and reviews how her description has been represented across different specification categories. She confirms that the prompt-derived items capture the intent she had in mind and clicks to continue  \circled{deepblue}{4}.
\begin{figure}[h]
    \centering
    \includegraphics[width=1\linewidth]{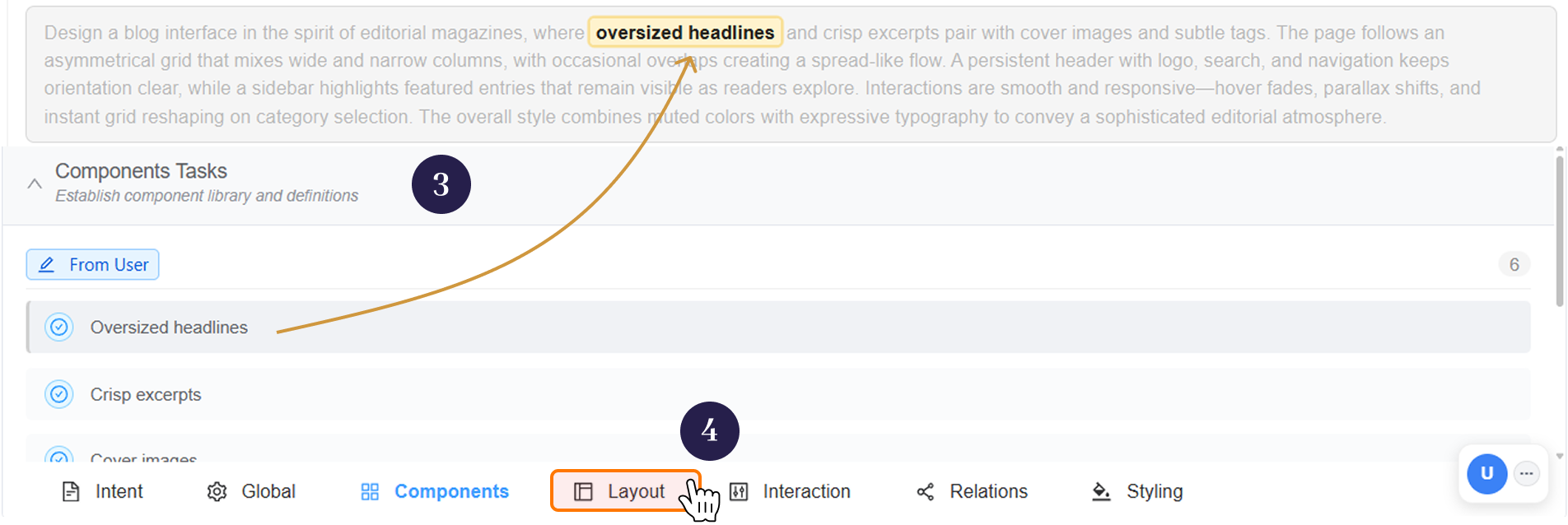}
    \label{fig:3-4}
\end{figure}

As Emma advances through the process, LegoUI presents newly inferred items in the Task Panel, each derived from the specification constructed so far  \circled{deepblue}{5}. Emma reviews each item and decides whether it should be retained in the design.

When reviewing an interaction-related entry stating “Article cards should support hover states to signal interactivity,” Emma recognizes this behavior as consistent with the editorial experience she envisions and retains the item as part of the specification.

When reviewing the styling-related entries, Emma encounters a suggestion stating, “The logo should use strong contrast to reinforce brand recognition.” She considers the scope of her workshop proposal and decides to center the interface on article content rather than brand identity. She marks the item as rejected and continues  \circled{deepblue}{6}.
\begin{figure}[h]
    \centering
    \includegraphics[width=1\linewidth]{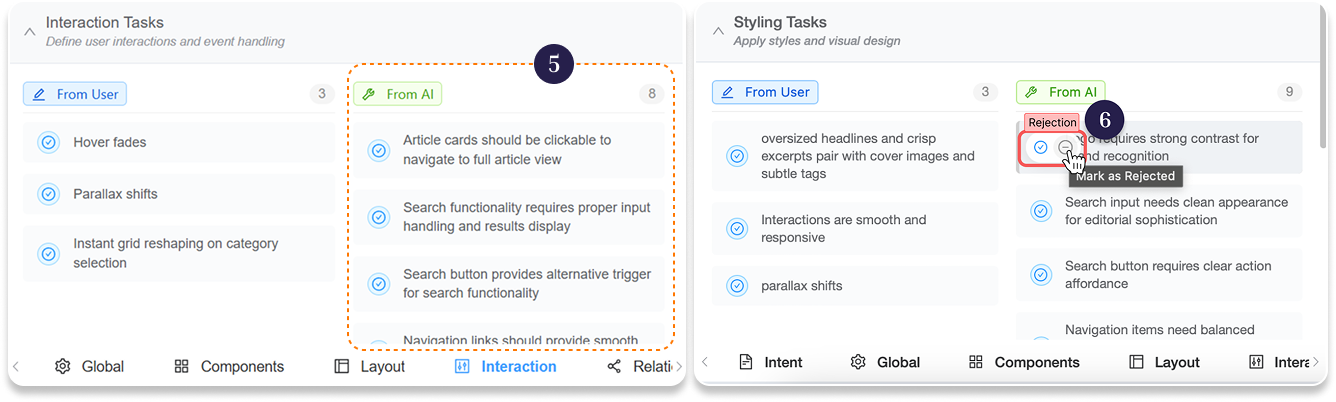}
    \label{fig:5-6}
\end{figure}

As Emma reads through the existing entries, she thinks about how editorial blogs often guide attention toward selected stories. She decides to make this intention explicit and adds a new item, writing: “Featured articles should be visually distinguished from the main feed.”  \circled{deepblue}{7}.
\begin{figure}[h]
    \centering
    \includegraphics[width=1\linewidth]{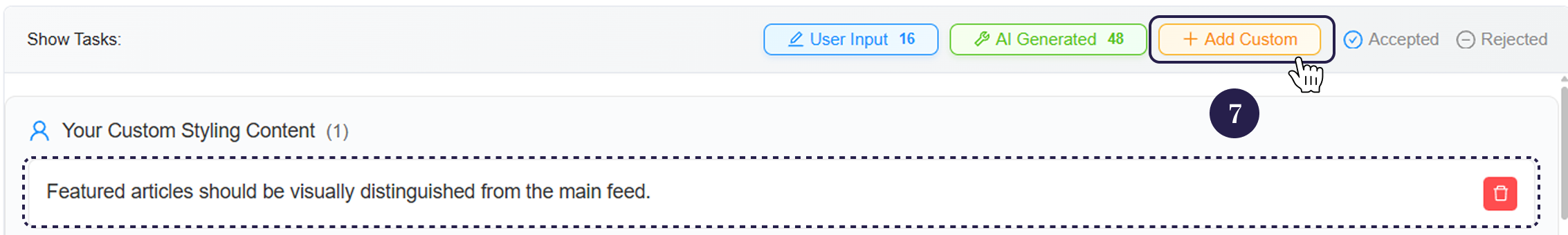}
    \label{fig:7}
\end{figure}

Once she is satisfied with the decisions captured so far, Emma clicks to move forward. She now sees a structured set of design elements ready to be used for interface generation.

\noindent\textbf{Generating an Initial Interface:}
With the specification in place, Emma clicks to generate an interface based on the current decisions  \circled{deepblue}{8}. LegoUI renders a blog layout that reflects the items she has retained.

\noindent\textbf{Refining and Exploring Further:}
After viewing the first generated interface, Emma decides to refine how articles are emphasized on the page. The initial version centers on a single lead story (Version 1). She revises the specification to adjust this emphasis and regenerates the interface, producing a version where articles are presented with more balanced visual weight (Version 2).

Emma explores a different direction by revisiting an earlier layout decision. She switches to a grid-based arrangement and generates another version, where articles are displayed in a compact grid that supports visual scanning (Version 3). 
Emma can switch between generated versions and select the one she prefers as the basis for further exploration. This process supports design ideation by enabling co-design through comparison and selection of alternatives.

\begin{figure}[h]
    \centering
    \includegraphics[width=1\linewidth]{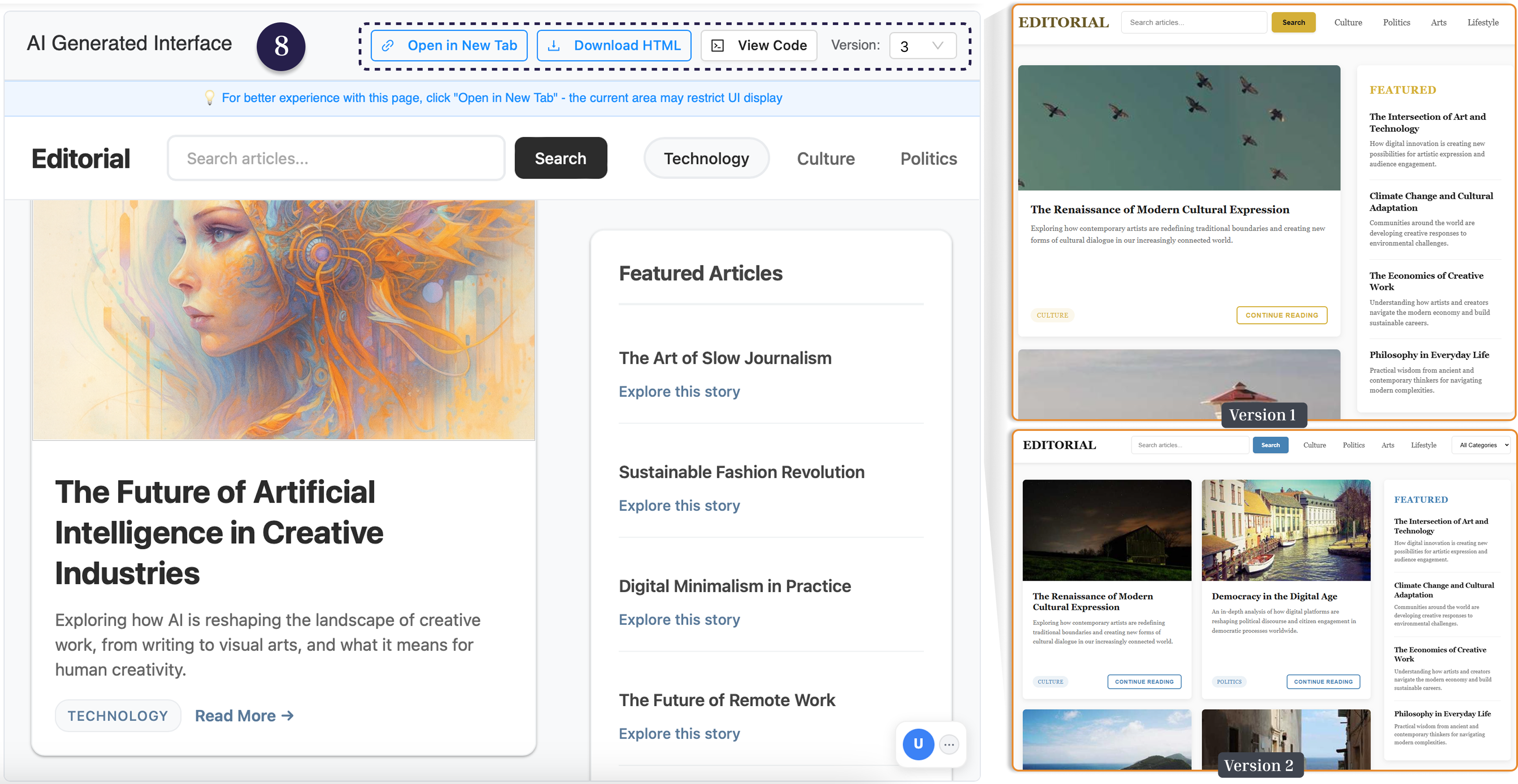}
    \label{fig:8}
\end{figure}
\section{UI-DSL as an Intermediate Design Representation}

Generative UI design involves a series of decisions about intent, content, structure, behavior, and visual presentation. In prompt-to-UI generation, many of these decisions are made implicitly by the model as it interprets the user's prompt and fills in missing details. When these decisions are represented only in the final interface or generated code, users have limited opportunities to inspect how their intent has been interpreted before the interface is produced.

\tool{} uses a UI-specific domain-specific language, or UI-DSL, to externalize these intermediate design decisions. The UI-DSL records the evolving design state during generation and provides a shared representation that both the model and the user can act on. Each recorded decision becomes an editable item that can be inspected, accepted, rejected, or revised as the interface is constructed. The full grammar and representative examples are provided in Appendix~\ref{sec:uidsl}.

\subsection{Externalizing Design Decisions}

The UI-DSL serves as the intermediate layer between a natural language prompt and the generated interface. It captures how the system interprets the prompt, how it extends underspecified requirements, and how these decisions accumulate across stages. This representation allows \tool{} to preserve earlier design decisions while continuing to refine later ones.

This structure is useful because UI generation often requires the model to infer information that users have left open. A prompt may describe the intended experience or visual direction while leaving the exact components, layout, navigation behavior, or styling details unspecified. The UI-DSL makes these inferred decisions visible before they become embedded in the output. Users can then judge whether the system's interpretation matches their intent and revise the design state when needed.

The UI-DSL also supports continuity across iterations. Instead of regenerating an interface from the original prompt alone, \tool{} generates from the current specification state. This allows later outputs to reflect both the user's initial intent and the accumulated decisions made during the staged process.

\subsection{Organizing Decisions for Staged Inspection}

The UI-DSL organizes design decisions according to the dimensions used in the staged generation workflow. It records high-level intent and global context, the set of interface components, layout structure, interaction behavior, relations among interface elements, and style direction. Each dimension provides a point where the system can surface its interpretation and where users can inspect a specific type of decision.

This organization helps users reason about the design process in smaller steps. During early stages, users can examine whether the system has captured the main task goals and required components. Later stages surface decisions about layout, behavior, relations, and style. This staged structure allows users to focus on one aspect of the design at a time while maintaining a specification that connects these aspects across the whole interface.

The UI-DSL therefore provides more than a storage format for generated content. It defines the units of inspection and intervention in \tool{}. A component entry, layout item, interaction rule, relation, or style item can each be presented to the user as a design decision. These items become the basis for the accept, reject, and add operations used throughout the system.

\subsection{Supporting Revision Through Provenance}

Each UI-DSL item carries provenance information that records how the decision entered the design state. Provenance distinguishes decisions derived from the user's prompt, decisions inferred by the model, and decisions introduced or modified by the user. This allows users to understand whether an item reflects their original input, an AI-generated extension, or a later revision.

User actions update the active design state through these items. Accepting an item preserves it for later stages. Rejecting an item removes it from the assumptions used for subsequent generation. Adding a new item introduces an explicit user requirement into the specification. These operations allow user feedback to influence the reasoning process directly, since later stages generate from the revised specification.

Provenance also helps maintain a trace of how the design evolves. As the system moves through staged reasoning, each decision remains connected to its source and revision history. This trace supports transparency during generation and provides a basis for explaining why a later interface element appears in the output. In this way, the UI-DSL enables \tool{} to treat user input, model inference, and user revision as part of the same evolving design state.
\section{LegoUI: System Design}
\begin{figure*}[t]
    \centering
    \includegraphics[width=1\linewidth]{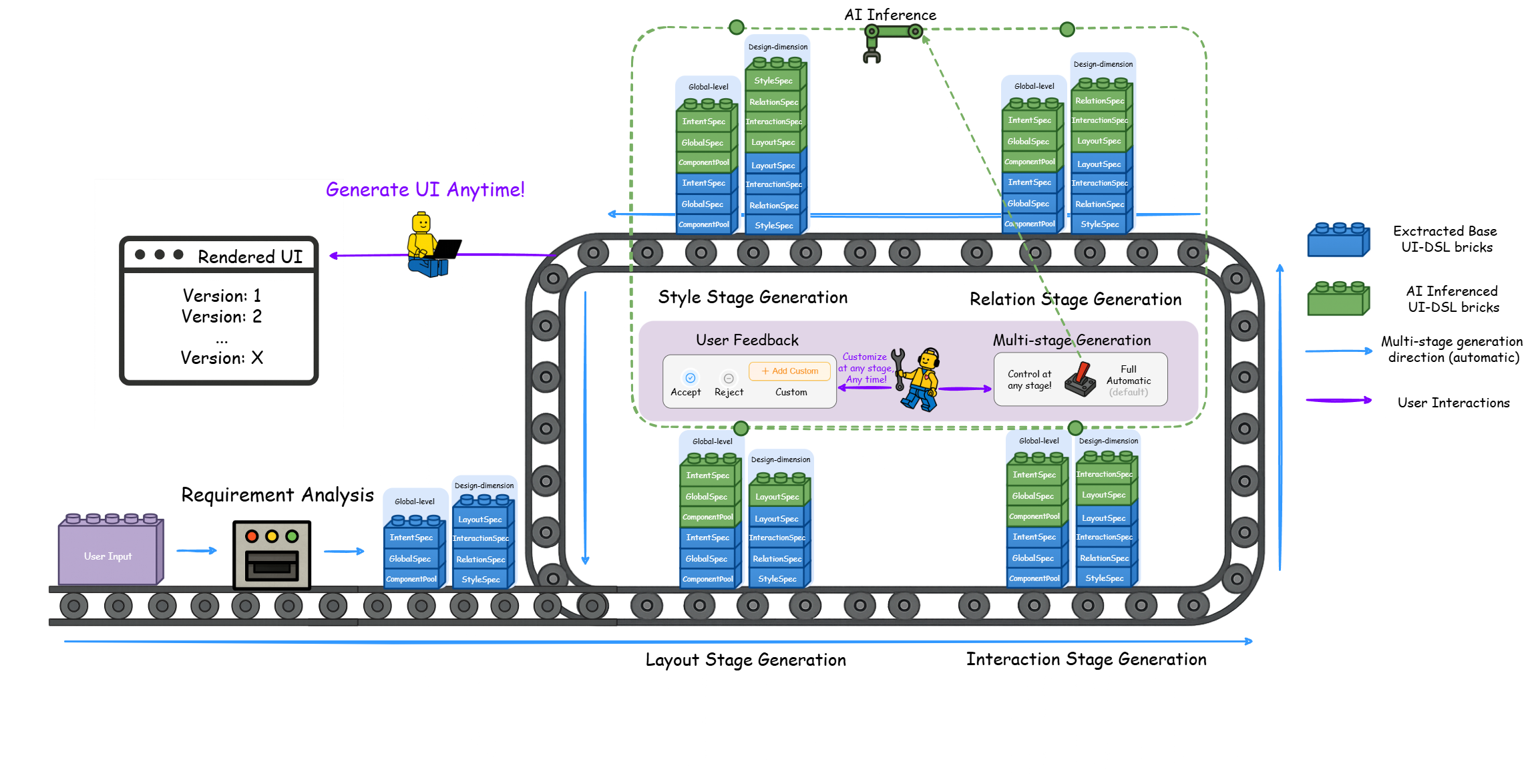}
    \caption{Overview of our pipeline for UI generation, illustrated through a factory-style production line metaphor. From a user input, the system performs \textit{requirement analysis} to extract \textit{Base UI-DSL} bricks, each embedding provenance for traceability. These bricks then progress along the conveyor through the \textit{Layout}, \textit{Interaction}, \textit{Relation}, and \textit{Style} Stage Generations, where robotic arms illustrate \textit{AI Inference} and workers symbolize \textit{User Feedback} operations such as \textit{accept}, \textit{reject}, \textit{add custom}, which users can apply at any stage of the pipeline. The evolving DSL can be rendered into UI outputs at any point, producing successive versions (\textit{Version 1}, \textit{Version 2}, …, X).}
    \label{fig:pipeline}
\end{figure*}

This section describes the system design of LegoUI and the process through which a UI-DSL specification is incrementally constructed from a user prompt. 
LegoUI treats each UI-DSL item as a manipulable unit, or brick, and organizes specification construction as an assembly process guided by staged reasoning. 
Figure~\ref{fig:pipeline} illustrates the system pipeline and shows how UI-DSL items move through successive reasoning stages while remaining open to user intervention during construction.

\subsection{Base UI-DSL Construction}
The construction process begins with forming a Base UI-DSL from the user prompt. 
This Base UI-DSL consists of UI-DSL items that capture early design interpretations expressed in the prompt and define the starting point for subsequent reasoning.

LegoUI derives these UI-DSL items by interpreting the prompt into structured statements that can be independently inspected and revised. 
Each item functions as a brick that represents a discrete design commitment and carries provenance linking it to the prompt. 
The resulting specification establishes a shared reasoning context that remains accessible throughout the construction process. The Base UI-DSL provides a stable basis for later reasoning stages, where additional UI-DSL items are introduced through further inference while remaining grounded in the existing specification.

\subsection{Dimension-wise Expansion}
After the Base UI-DSL is established, LegoUI expands the specification by incrementally writing new DSL items into dimension-level specifications. Each expansion step operates on the current UI-DSL state and appends inferred design decisions that extend the existing reasoning context.

One class of expansion introduces layout structure while refining the component pool and global specification. At this point, the UI-DSL already contains component definitions extracted from the prompt. Based on these definitions, the model may construct layout regions by binding components into page-level structure, and may introduce additional component attributes required to support the inferred arrangement. For instance, when a feed-style arrangement is inferred for a set of article cards, the system can write a layout region
\[
R_{\textit{Feed}}=(N_r,\{\mathit{dir}=\textit{vertical}\},\{\texttt{Comp}(C_{\textit{ArticleCard}})\},P_l)
\]
into $\textsc{LayoutSpec}$, while augmenting the corresponding component definition with attributes that support repeated presentation, such as scroll behavior or item spacing, recorded as new assertions in $A_c$ with associated provenance.

In addition to layout structure, expansion may introduce global requirements implied by the evolving specification. For example, certain structural configurations can lead to the addition of navigation routes written into $\textsc{GlobalSpec}$. These entries are represented as DSL items that share the same provenance structure and persist as part of the reasoning state.

All newly introduced DSL items are surfaced to the user as model-generated decisions, each linked to its provenance entry. User actions operate directly on these items. Accepted decisions remain active in the specification and serve as context for subsequent expansion, while revisions update the reasoning state consumed by later inference.

Other expansion classes introduce interaction rules that reference components already placed within the layout and attributes established in the component pool. For example, a navigation behavior may be expressed as
\[
\texttt{Click}_{\textit{ArticleCard}}[\texttt{PageIs}(\textit{Feed})]\Rightarrow\{\texttt{NavigateTo}(\textit{ArticlePage})\}\langle P_i\rangle,
\]
which is appended to $\textsc{InteractionSpec}$ and linked to existing layout and component identifiers. Relation and style expansions follow the same pattern, with each DSL item written against the accumulated UI-DSL and recorded as an extension of the current reasoning state.

\subsection{User-guided Reasoning Correction}

During UI-DSL construction, users participate in the reasoning process by updating the provenance of DSL items. Each generated item is initially active in the current reasoning state, and user edits append provenance entries that influence how subsequent inference interprets specific design decisions.

When a user rejects a DSL item, the system records a provenance entry with source $\texttt{user\_edit}$ and subtype $\texttt{remove}$. This entry marks the decision as excluded from the active reasoning state. Subsequent expansion steps read this signal and constrain inference to avoid deriving decisions conditioned on it.

When a user introduces a new requirement, the system creates a new DSL item with a provenance entry of subtype $\texttt{add}$. This entry records the injected intent as an explicit assumption and incorporates it into the reasoning context consumed by later expansion stages.

Because user edits are represented as provenance updates, their effects propagate uniformly across specifications. The reasoning state at any point reflects the cumulative interpretation of provenance entries. An excluded item may later receive a provenance entry indicating acceptance, allowing the reasoning state to evolve while preserving a complete trace of prior decisions.






\subsection{UI Rendering}
After construction, the UI-DSL represents a complete design state that incorporates user edits. The model generates a UI instance by consuming the current UI-DSL and translating the recorded design decisions into HTML-based interface code. The generated interface corresponds to the specification state at the moment of generation.

After an initial UI instance is produced, the UI-DSL remains available for further updates through the same operations used during construction. Users may revise the state of DSL items, after which subsequent generation consumes a user-selected UI-DSL state to produce a new UI instance or a variation derived from that state.

\subsection{Implementation}
LegoUI is implemented as a web-based system, with a frontend built using Vue3 and Ant Design and a backend implemented with Python and FastAPI. The backend invokes large language models through structured prompts, using zero-shot prompting with fragmentary examples for grammar variables, with Claude Sonnet4 as the primary model and support for GPT-4o and Gemini 2.5.
The rendered interface is generated as HTML code from the current UI-DSL state and refreshed whenever a new generation is triggered.

\section{Technical Evaluation}
To assess how effectively \tool{} supports requirement analysis, we conducted a technical evaluation focusing on the system’s ability to decompose design prompts into structured representations of user requirements. The goal of this evaluation is to determine whether the system can faithfully capture key requirements under a controlled complexity setting, thereby establishing a foundation for subsequent stages of the design pipeline (e.g., layout generation and interactive prototyping). 
Specifically, we evaluated the generated task checklists
along three dimensions:
\begin{itemize}
    \item \textbf{Prompt Structuring Accuracy} – the extent to which the system correctly rewrites and organizes prompt content into a structured checklist.
    \item \textbf{UI Content Coverage} – the extent to which all seven dimensions of the specification (intent, global, component, layout, interaction, relation, and style) are represented from the prompt.
    \item \textbf{Semantic Fidelity} – the extent to which the extracted checklist preserves the original design intent without omissions, distortions, or misinterpretations.
\end{itemize}

\subsection{Setup}
\subsubsection{Dataset}
Firstly, we collected 36 long-form UI design prompts from the UIPrompt\footnote{\url{https://uiprompt.art/}} website. The original prompts vary substantially in length, detail, and UI content density. To ensure comparability, we standardized them using GPT-4o to produce rewritten versions that meet fixed complexity criteria.
\begin{itemize}
    \item Length: 150–200 words (~110–150 tokens) per prompt.
    \item Component Count: 6–10 distinct UI components.
    \item Interaction Types: 2–4 distinct interactions (e.g., click, hover, form validation).
    \item Stylistic Constraints: 1–3 explicit style requirements (e.g., color theme, typography, alignment).
\end{itemize}

To further balance the corpus, after verifying that the 36 ground-truth prompts contained no \orange{relationSpec}. We then added four GPT-4o–generated prompts of matched length and complexity, yielding a dataset of 40 prompts.
To illustrate the type of standardized prompts used in the evaluation, consider the following excerpt:

\begin{quote}
\textit{``Design an image gallery experience inspired by an artist’s sketchbook. Use high-detail paper textures and subdued paper tones; attach photos with corner mounts, tape or clips. Include realistic page-turning and zoom interactions, hand-written annotations and loading animations that look like drawings. Navigation tabs should resemble bookmarks or sticky notes, filtering controls like hand-labeled dividers, so that browsing feels like flipping through a cherished art portfolio.''}
\end{quote}

This example states explicit requirements across component specification, interaction behavior, and style constraints, while remaining within the controlled length and complexity bounds. Other prompts in the dataset vary in application domain (e.g., media, dashboards, productivity) but follow the same standardization procedure.

\subsubsection{Ground Truth Construction}

For each standardized prompt, evaluators used our system to perform the requirement analysis step, which generates a \purple{PromptChecklist} organized according to the system’s top-level DSL structure: 
\orange{intentSpec}, \orange{globalSpec}, \orange{componentPool}, \orange{layoutSpec}, \orange{interactionSpec}, \orange{relationSpec}, \orange{styleSpec}
At this stage, we extract only explicitly stated information, including elements shown directly or mentioned in the prompt text.

For each prompt, two independent annotators created a reference \purple{PromptChecklist} containing all explicit items. Annotators worked independently and then resolved disagreements through discussion. Prior to annotation, they were trained on example prompts to align their assessments. This ground truth (GT) served as the reference standard for evaluation.

\subsubsection{Scoring Procedure}
To systematically compare system outputs with the reference checklists, we adopted a unit-level coding scheme. Each item extracted by the system was aligned against the corresponding ground truth entry and assigned a categorical label. The scheme distinguishes between correctly captured items, partially aligned items, and various forms of error. In particular, we separate content errors (where the semantic unit itself is incorrect) from placement errors (where the semantic unit is valid but assigned to the wrong specification category). This separation allows us to diagnose whether failures arise primarily from semantic misunderstanding or structural misalignment. The full coding categories are listed in \autoref{tab:definitions_of_codes}.

\begin{table}[t]
    \centering
    \caption{Definitions of annotation codes used for evaluating extraction quality.}
    \setlength{\tabcolsep}{4pt}
    \begin{tabularx}{\columnwidth}{l|l|Y}
    \hline
    \textbf{Code} & \textbf{Name} & \textbf{Definition} \\ \hline
    MT   & Matched              & Fully matches the GT entry in content and placement. \\ \hline
    PM   & Partial Match        & Matches GT in intent but omits details or attributes. \\ \hline
    ML   & Misaligned           & Semantically correct but assigned to the wrong DSL category. \\ \hline
    OM   & Omitted              & GT item not extracted by the system. \\ \hline
    EX   & Extraneous           & Not in GT and not explicitly stated in the prompt. \\ \hline
    \end{tabularx}
    \label{tab:definitions_of_codes}
    
    \vspace{0.5mm}
    \begin{flushleft}
    \footnotesize{
    \textit{Metric Computation.} \\
    Accuracy = (MT + PM) / (MT + PM + ML + EX); 
    Coverage = (MT + PM) / (MT + PM + OM); 
    Redundancy Rate = EX / (MT + PM + ML + EX). \\
    Partial Matches (PM) are included in accuracy since they capture the correct semantic unit even if attributes are missing. 
    Redundancy rate follows over-generation measures in text generation.}
    \end{flushleft}
\end{table}

\subsubsection{Coding Process}
Two coders with expertise in UI design and interface specification independently applied the coding scheme to all system outputs. They examined each generated checklist entry, compared it with the ground truth, and assigned the appropriate label. To ensure reliability, discrepancies between coders were determined by a third reviewer. The final metrics were calculated by aggregating the coded results in all prompts and checklist categories.

\subsection{Results}
\begin{table}[ht]
\centering
\caption{Section-level Coding Results (percentages of coded items) }

\label{tab:section_coding}
\begin{tabular}{lcccccccc}
\toprule
Section & Matched (\%) & Partial Match (\%)  & Misaligned (\%) & Omitted (\%) & Extraneous (\%) & N \\
\midrule
intentSpec         & 95.95 & 1.35 &  2.70 & 0.00 & 0.00 & 74 \\
globalSpec         & 91.67 & 0.00 & 7.14 & 1.19 & 0.00 & 84 \\
componentPool & 96.36 & 0.00 &  2.42 & 1.21 & 0.00 & 165 \\
layoutSpec         & 86.96 & 8.70 &  4.35 & 0.00 & 0.00 & 23 \\
interactionSpec    & 93.65 & 1.59 &  4.76 & 0.00 & 0.00 &  63 \\
relationSpec       & 100.00 & 0.00 &  0.00 & 0.00 & 0.00 & 14 \\
styleSpec        & 80.33 & 1.64 &  9.84 & 8.20 & 0.00 & 61 \\
\bottomrule
\end{tabular}
\end{table}

\begin{table}[ht]
\centering
\caption{Section-level Accuracy, Coverage, and Redundancy}
\label{tab:section_metrics}
\begin{tabular}{lccc}
\toprule
Section & Accuracy (\%) & Coverage (\%) & Redundancy (\%) \\
\midrule
intentSpec         & 97.30 & 100.00 & 0.00 \\
globalSpec         & 92.77 & 98.72 & 0.00 \\
componentPool & 97.55 & 98.76 & 0.00 \\
layoutSpec         & 95.65 & 100.00 & 0.00 \\
interactionSpec    & 95.24 & 100.00 & 0.00 \\
relationSpec       & 100.00 & 100.00 & 0.00 \\
styleSpec        & 89.29 & 90.91 & 0.00 \\
\bottomrule
\end{tabular}
\end{table}

Across the dataset, the ground truth comprised 484 entries from 40 prompts, averaging 9.7 items per prompt (median = 11). The distribution was uneven across sections. Component Pool accounted for over one-third of all entries, while globalSpec, Intent, and interactionSpec each contributed between 13–17\%. By contrast, layoutSpec and relationSpec made up less than 8\%. On average, each prompt included several component entries ($\approx$3.4) but fewer than one entry in layout or relation.

Most categories achieved high match rates, as shown in Table~\ref{tab:section_coding}. This confirms that the requirement analysis stage effectively structures explicit prompt content into a machine-checkable form. Component Pool and Intent both maintained accuracies above 95\%, showing that the system reliably captures atomic UI elements and overarching design goals necessary for downstream generation. interactionSpec also performed well, with nearly complete coverage, and both layoutSpec and relationSpec achieved perfect scores, indicating that spatial and relational information expressed in the prompts was consistently extracted without error.

styleSpec showed the weakest performance, with only 80.33\% of items matched and 8.20\% omitted. Many of these errors arose from ambiguity in the boundary between styleSpec and other category, such as whether “minimalist alignment” should be categorized as layout or style, or whether “dark mode” should be treated as a global constraint or a visual directive. These patterns reflect two recurring forms of confusion: globalSpec versus styleSpec, and layoutSpec or interactionSpec versus styleSpec. Omitted items were also concentrated in visual properties such as color choices and animation details, which were less consistently extracted when not explicitly enumerated. By contrast, layoutSpec and relationSpec both reached perfect scores, but the number of explicit items in these sections was noticeably lower in the collected prompts. 

This may reflect a characteristic of open-ended UI generation, where users give free-form descriptions of desired components, interactions, and visual design, but provide fewer details about component relationships or precise layout requirements.

The metrics in Table~\ref{tab:section_metrics} reinforce these observations. Accuracy exceeded 95\% in almost all sections. Coverage was also close to complete, and redundancy remained at 0\%. Several sections even achieved perfect scores, showing that the structured checklist specification can reliably capture explicit requirements. At the same time, styleSpec and the alignment of global constraints emerged as areas that would benefit from refinement of extraction rules or additional validation mechanisms.

\section{User Study}
We conducted a user study to investigate how \tool{}, our scaffolded generation pipeline, reshapes the process of creating and refining user interfaces. The primary goal of this study was to assess the design goals detailed in Section~\ref{sec:3_DG}, specifically focusing on transparency (DG1), control of the interface (DG2), and the ability of quick customization as desired (DG3). 

To evaluate these design goals, we formulated a series of research questions (RQs) to assess \tool{}'s capabilities:

\textbf{RQ1.} Does~\tool{} produce user interfaces that participants perceive as more aligned with their design goals? How do these evaluations compare to those of existing one-shot systems (baseline systems)?

\textbf{RQ2.} How does the presence of intermediate stages influence users’ perceived control, clarity, and fairness when shaping the generated UIs?

\textbf{RQ3.} How do editing practices differ between~\tool{} and one-shot generated UI systems conditions? What are the resulting differences in workload, strategies, and satisfaction?

\textbf{RQ4.} What challenges and tensions arise as participants adapt generated UIs to their needs? To what extent can these be attributed to the limits of the pipeline itself versus the broader interaction paradigm?

\subsection{Participants}
We recruited 15 participants (8 female and 7 male, aged under 24 to early 30s). The group included graduate students, researchers, and developers. As summarized in Table~\ref{tab:participant_data}, most participants had limited prior experience with generative design tools, with 11 reporting beginner-level familiarity and four reporting intermediate familiarity. Prior to the experiment, participants were provided with the Participant Information Statement and Consent Form by email and were given time to consider their participation. Before participation, participants were informed about the purpose of the study, the study procedure, the voluntary nature of their participation, their right to withdraw at any time, and how their data would be used. Verbal informed consent was obtained from all participants before any data-generating activities. The study was approved by our university's Institutional Review Board (IRB) to ensure adherence to ethical research standards, and all participants were compensated with approximately AUD \$30 for their time.

\paragraph{Ethics approval.}
This study was approved by the Human Research Ethics Committee at the University of New South Wales (UNSW), reference number iRECS9443. The study was reviewed as a low-risk human ethics application.

\begin{table*}[h]
    \centering
    \caption{Demographics of participants.}
    \resizebox{0.9\linewidth}{!}{%
    \begin{tabular}{l|l|l|l|l|l}
        \hline
        \textbf{ID} & \textbf{Age} & \textbf{Professional Background} & \textbf{GenDesign Tool Experience} & \textbf{GenDesign Tool Use} & \textbf{GenAI Tool Use}\\
        \hline
        P1  & 30s      & Researcher       & Intermediate & Moderate & Frequent \\
        P2  & Under 24 & Graduate Student    & Beginner & Rare & Moderate \\
        P3  & 25-30    & Researcher       & Beginner & Rare & Moderate\\
        P4  & 25-30    & Developer        & Beginner     & Rare  & Frequent\\
        P5  & 25-30    & Researcher       & Intermediate & Moderate & Frequent\\
        P6  & 25-30    & Researcher       & Beginner & Rare & Frequent\\
        P7  & 25-30    & Graduate Student    & Beginner & rare & Moderate\\
        P8  & Under 24 & Graduate Student    & Beginner & Rare & Moderate\\
        P9  & 25-30    & Graduate Student    & Beginner & Rare &Moderate\\
        P10 & 25-30    & Developer        & Beginner & Rare & Frequent\\
        P11 & 30s      & Developer        & Intermediate & Frequent & Frequent\\
        P12 & 25-30    & Developer        & Beginner     & Rare & Frequent\\
        P13 & 25-30    & Researcher       & Intermediate & Moderate & Frequent\\
        P14 & 25-30    & Graduate Student    & Beginner & Rare & Moderate\\
        P15 & Under 24 & Graduate Student & Beginner & Rare & Rare\\
        \hline
    \end{tabular}
    }%
    \label{tab:participant_data}
\end{table*}

\subsection{Study Design}\label{subsec: task_design}

We employed a within-subject design. Specifically, the following generate UI tasks were examined:

\paragraph{Task 1: Multi-Tool UI Output Evaluation}
We selected three representative prompts from an earlier technical evaluation, covering common design scenarios: a dashboard, a portfolio page, and a productivity-oriented note-taking interface. For each prompt, we generated five UI results from five sources: our \tool{}, the Claude model, and three industry tools (Bolt, Lovable, and Vercel v0). These UIs were presented to participants one-by-one in a randomized and anonymized order. Participants rated each UI on a 5-point Likert scale across five quality dimensions: Intent Alignment, Usability, Accessibility, Visual Coherence, and Trustworthiness. Additionally, they were asked to rank their top two preferred outputs for each prompt. The highest-scoring industry tool (Bolt) was selected for subsequent tasks.

\paragraph{Task 2: System Interaction}
Based on Task 1, we selected Bolt as the representative of industry tools. Participants were asked to use our \tool{}, Claude Chat and Bolt to complete a full prompt-to-UI workflow. Appendix~\ref{A: example} illustrates how users can interact with ~\tool{}. The~\tool{} has 5 stages, including requirement analysis, layout generation, interaction generation, relation generation and style generation. Participant needed to interact with each stage one-by-one. To ensure the task comparability, we prepared three prompts of similar length and complexity, all specifying the same overall purpose but differing in design details and functional requirements. Participants were divided into three groups, with each group assigned one of the prompts. The system guided them through sequential stages of generation, and the task captured how participants engaged with the process of shaping the UI. After completing the workflow, participants filled out a short questionnaire assessing their perceived control, clarity, and satisfaction.

\paragraph{Task 3: Post-Generation Editing}
Participants performed interactive editing tasks across three systems: \tool{}, Claude Chat, and Bolt. We also prepared three prompts of similar length and complexity but with distinct design specification. Each system condition was paired with a distinct prompt, and participants were divided into three groups so that cross the task all prompts were evenly distributed across systems.

For \tool{}, participants were further subdivided with editing tasks distributed across five fixed targets (Component, Layout, Relation, Interaction, and Style) to ensure balanced coverage of prompts. In the Claude and Bolt, participants worked on UIs that had already been generated in a one-shot pass from the assigned prompt, and then refined them through at least two rounds of modifications. After completing the editing task in each system, participants filled out a short questionnaire assessing their perceived control, fairness, workload, and satisfaction.

\subsection{Procedure}
At the start of the experiment, participants received a brief overview of the study and were provided with an explanation and a working example of each tool and interface. The experiment followed a within-subjects design with three distinct tasks. Before each task, participants were shown a tutorial video for reference. They were required to complete the tasks in a sequential order: Task 1, Task 2, and Task 3. After completing each task, participants filled out questionnaires related to the task, as detailed in Section \ref{subsec: task_design}.

Upon finishing all tasks, a semi-structured interview was conducted with each participant to gather reflections on their overall experience with the systems. On average, the study lasted approximately 75 minutes per participant. All sessions were conducted in person, with the exception of one interview that was held remotely via Zoom and voice-recorded.

\subsection{Results}
In this section, we present our findings from our user studies, evaluating the \tool{}’s capabilities. The results are summarized from the questionnaires of each task and the semi-structured interview conducted with participants. All 15 participants completed the experiment, and the full dataset was successfully collected and retained for analysis. 

We started our analysis by computing mean scores and standard deviations for each questionnaire item and dimension. To examine pairwise differences between systems, we conducted paired t-tests and report p-values with significance markers. We also report effect sizes (Cohen’s d) to indicate the magnitude of observed differences beyond significance testing.

\subsubsection{Task 1 Results: Multi-Tool UI Output Evaluation}
\begin{table*}[h]
\centering
\caption{%
Tallies of each system across three prompts (Dashboard, Productivity, Portfolio) and their cross-prompt averages. 
The five evaluation dimensions are: IA = Intent Alignment, U = Usability, A = Accessibility, VC = Visual Coherence, T = Trustworthiness. 
All scores are on a 5-point Likert scale (maximum = 5). 
Blue bold numbers indicate the highest score in each dimension, and underlined numbers indicate the second-highest score.%
}
\label{tab:UI_scores}
\resizebox{\textwidth}{!}{
\begin{tabular}{l|ccccc|ccccc|ccccc|ccccc}
\toprule
\multirow{2}{*}{\textbf{System}} 
 & \multicolumn{5}{c|}{Dashboard} 
 & \multicolumn{5}{c|}{Productivity} 
 & \multicolumn{5}{c|}{Portfolio} 
 & \multicolumn{5}{c}{Average across Prompts} \\
\cmidrule(lr){2-6} \cmidrule(lr){7-11} \cmidrule(lr){12-16} \cmidrule(lr){17-21}
 & IA & U & A & VC & T 
 & IA & U & A & VC & T 
 & IA & U & A & VC & T 
 & IA & U & A & VC & T \\
\midrule
Lovable     
 & 2.67 & 2.73 & 2.67 & 2.67 & 2.27  
 & 2.67 & 1.87 & 2.33 & 2.33 & 2.20  
 & \underline{4.13} & \underline{3.73} & \underline{3.73} & \underline{4.13} & \underline{3.93}  
 & 3.16 & 2.78 & 2.91 & 3.04 & 2.80 \\
Bolt        
 & 3.07 & 2.73 & 3.40 & 3.60 & 3.33  
 & 3.00 & 3.33 & 3.73 & 3.40 & 3.07  
 & 3.93 & 3.53 & 3.67 & 3.40 & 3.47  
 & 3.33 & 3.20 & 3.60 & 3.47 & 3.29 \\
Vercel V0   
 & 3.00 & 3.27 & 3.60 & 3.33 & 2.87  
 & \underline{3.73} & 3.67 & 3.73 & 3.33 & 3.27  
 & 3.27 & 3.47 & 3.33 & 3.27 & 3.13  
 & 3.33 & 3.47 & 3.56 & 3.31 & 3.09 \\
Claude Chat 
 & \underline{4.00} & \underline{3.67} & \underline{4.00} & \underline{4.40} & \underline{3.80}  
 & \underline{3.73} & \underline{4.07} & \underline{4.20} & \textbf{\textcolor{blue}{4.47}} & \underline{4.07}  
 & 3.73 & 3.47 & 3.40 & 3.73 & 3.47  
 & \underline{3.82} & \underline{3.73} & \underline{3.87} & \underline{4.20} & \underline{3.78} \\
\tool{} 
 & \textbf{\textcolor{blue}{4.27}} & \textbf{\textcolor{blue}{4.07}} & \textbf{\textcolor{blue}{4.40}} & \textbf{\textcolor{blue}{4.67}} & \textbf{\textcolor{blue}{4.47}}  
 & \textbf{\textcolor{blue}{4.20}} & \textbf{\textcolor{blue}{4.47}} & \textbf{\textcolor{blue}{4.53}} & \underline{4.40} & \textbf{\textcolor{blue}{4.27}}  
 & \textbf{\textcolor{blue}{4.20}} & \textbf{\textcolor{blue}{4.40}} & \textbf{\textcolor{blue}{4.53}} & \textbf{\textcolor{blue}{4.53}} & \textbf{\textcolor{blue}{4.53}}  
 & \textbf{\textcolor{blue}{4.22}} & \textbf{\textcolor{blue}{4.31}} & \textbf{\textcolor{blue}{4.49}} & \textbf{\textcolor{blue}{4.53}} & \textbf{\textcolor{blue}{4.42}} \\
\bottomrule
\end{tabular}
}
\end{table*}

\begin{figure}
    \centering
    \includegraphics[width=1\linewidth]{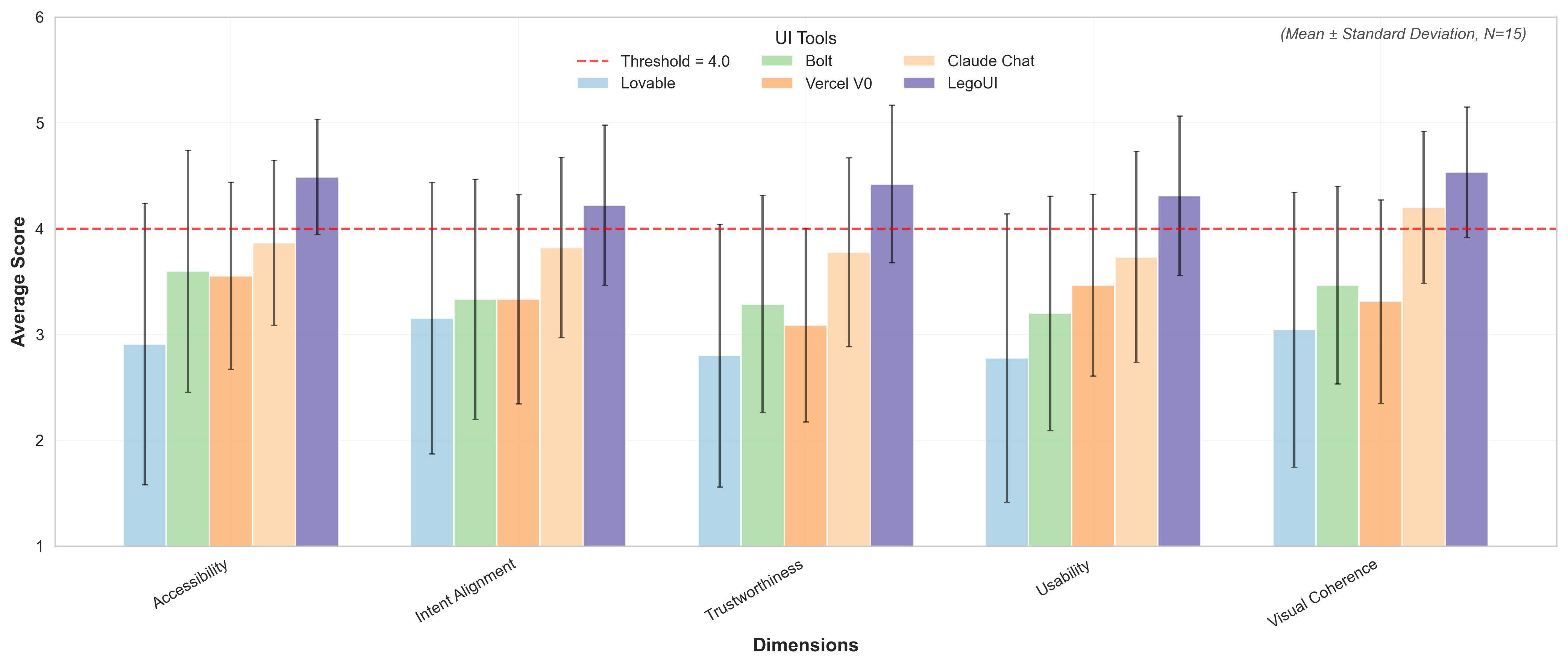}
    \caption{Average ratings of anonymized UIs across five evaluation dimensions (Accessibility, Intent Alignment, Trustworthiness, Usability, and Visual Coherence). Bars show mean values with standard deviations ($N=15$). The red dashed line indicates the threshold score of 4.0.}
    \label{fig:UI_Dimensions_Comparison}
\end{figure}

\begin{figure}[ht]
  \centering

  \begin{subfigure}{0.32\linewidth}
    \centering
    \includegraphics[width=\linewidth]{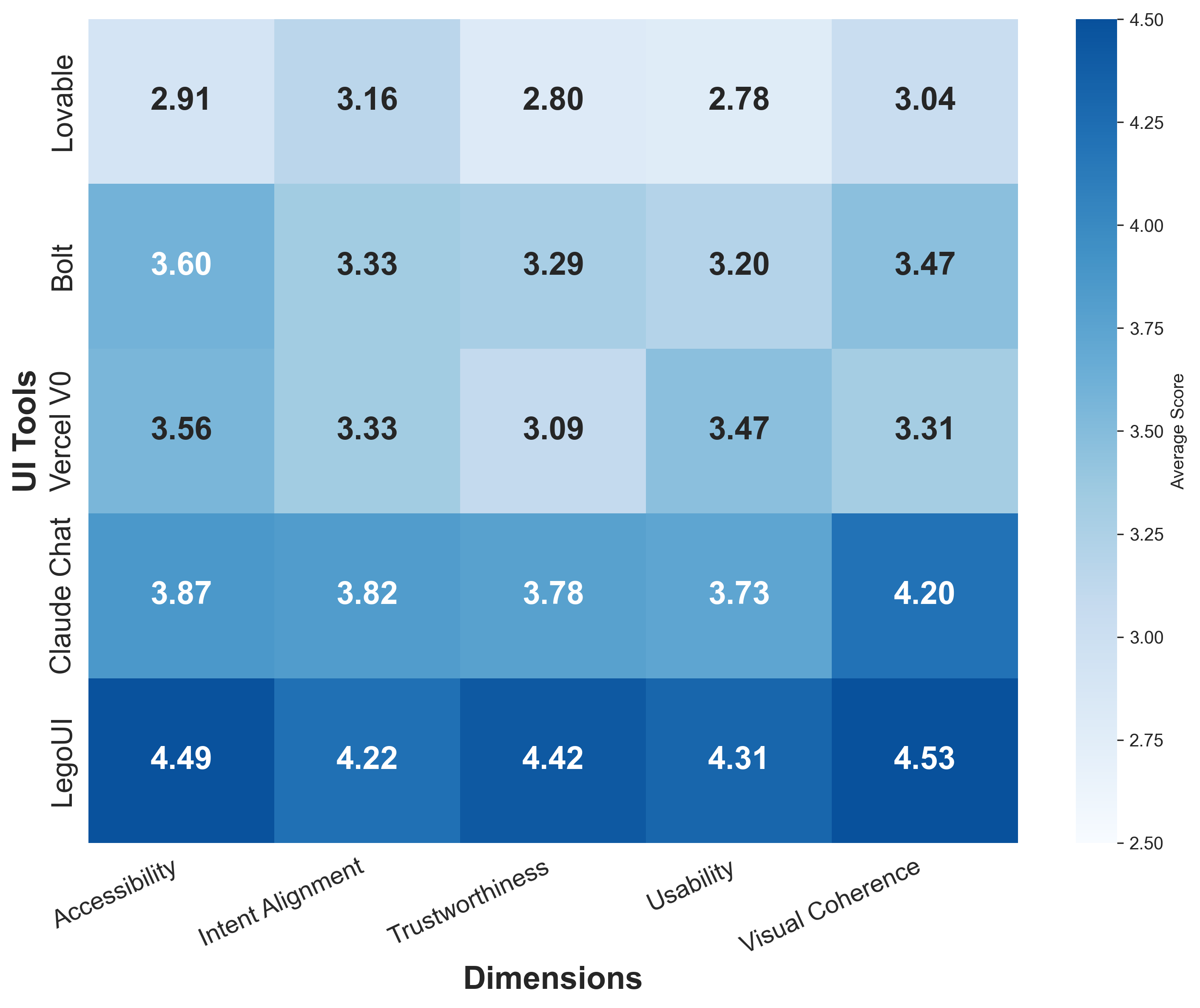}
    \caption{Average scores by dimension}
    \label{fig:ui_avg_scores}
  \end{subfigure}
  \hfill
  \begin{subfigure}{0.32\linewidth}
    \centering
    \includegraphics[width=\linewidth]{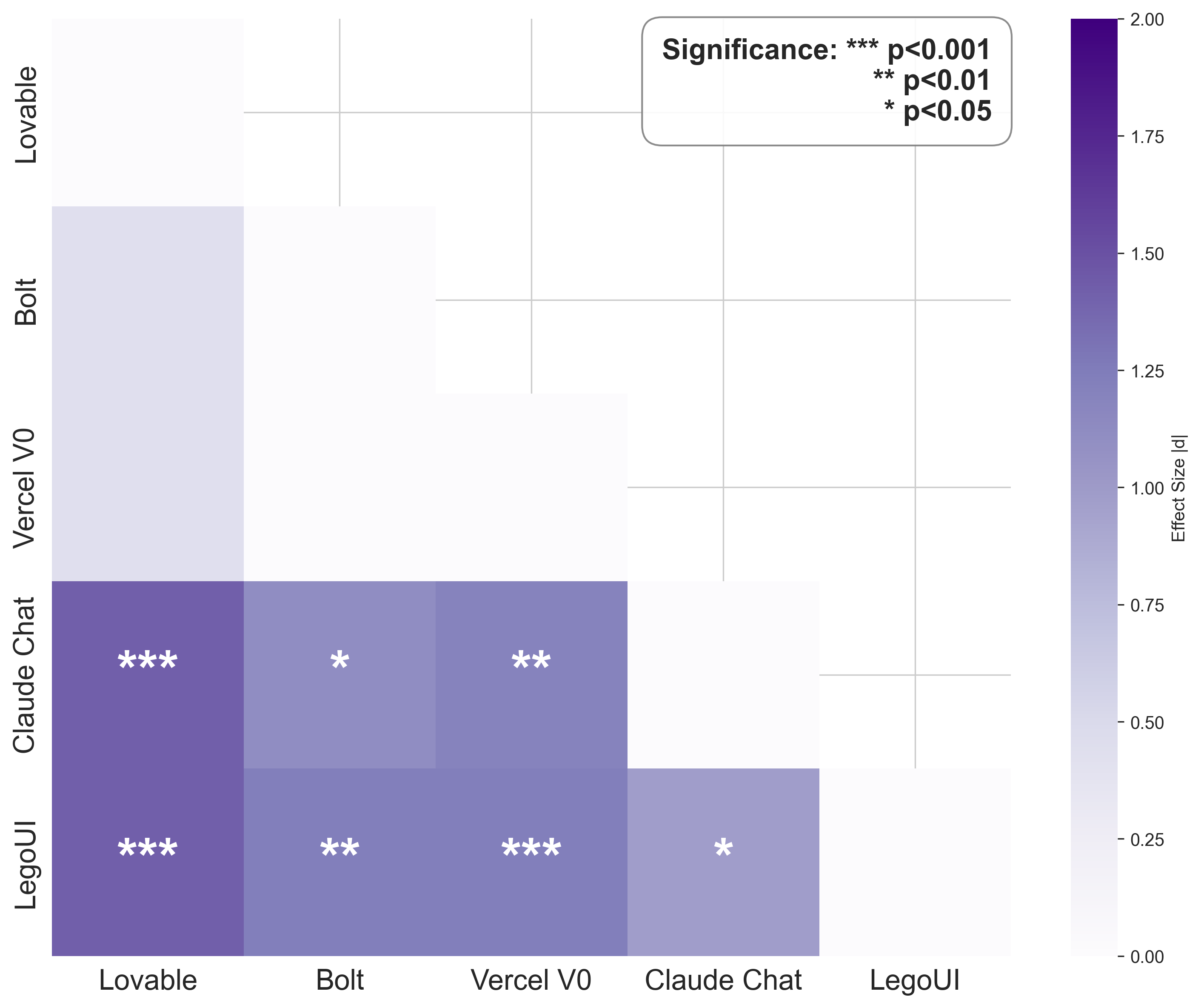}
    \caption{Intent Alignment}
    \label{fig:heatmap_intent}
  \end{subfigure}
  \hfill
  \begin{subfigure}{0.32\linewidth}
    \centering
    \includegraphics[width=\linewidth]{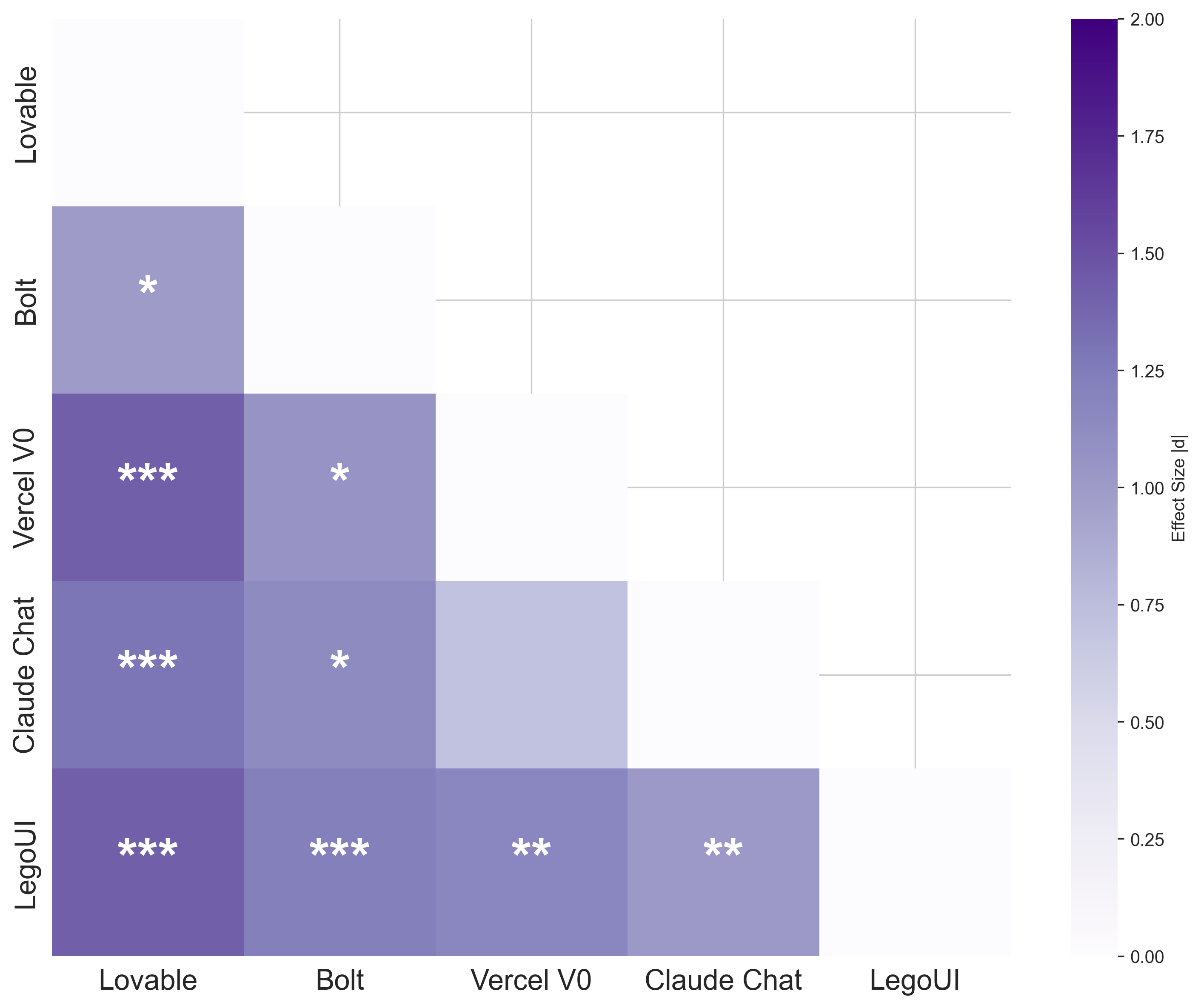}
    \caption{Usability}
    \label{fig:heatmap_usability}
  \end{subfigure}

  \begin{subfigure}{0.32\linewidth}
    \centering
    \includegraphics[width=\linewidth]{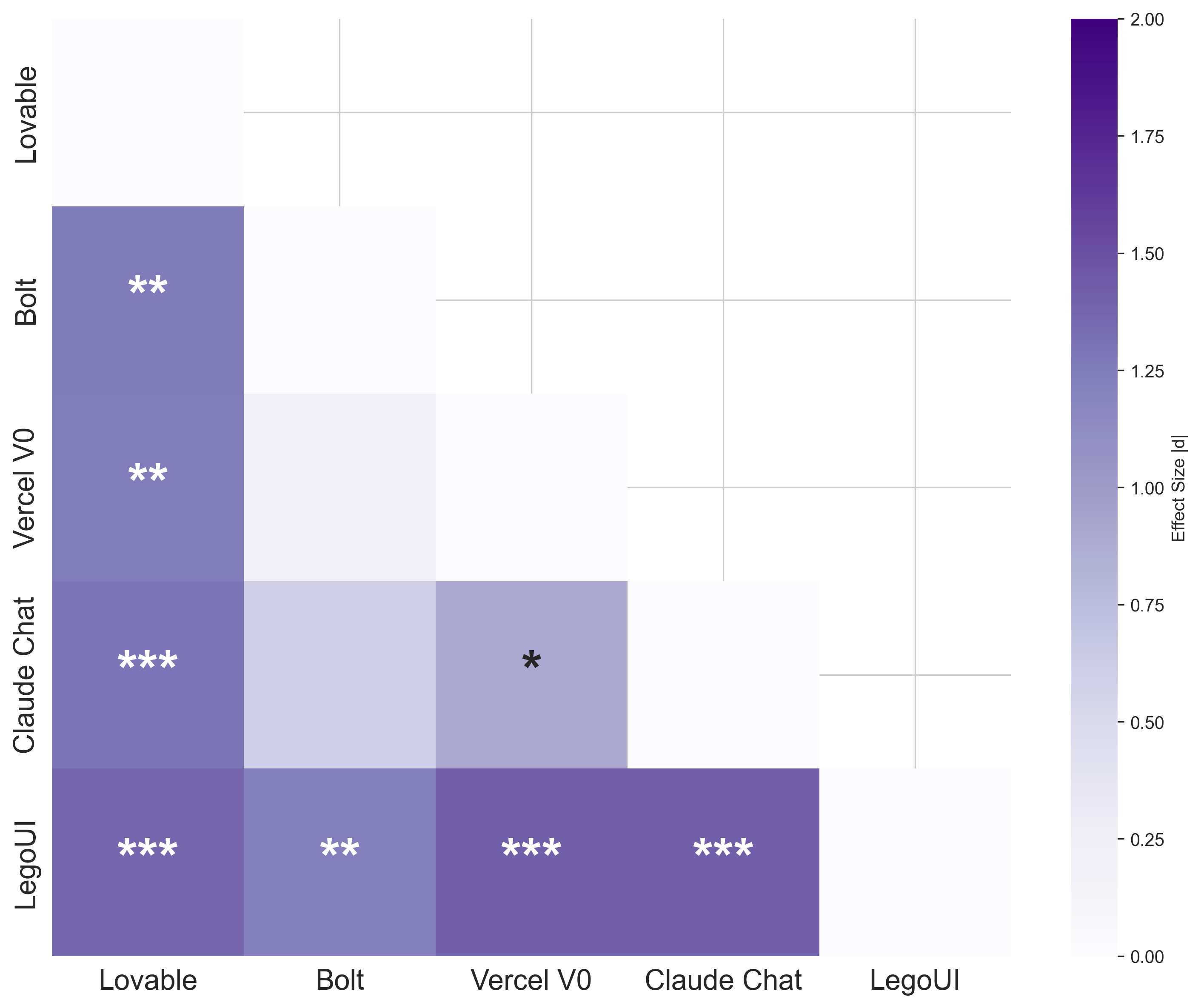}
    \caption{Accessibility}
    \label{fig:heatmap_accessibility}
  \end{subfigure}
  \hfill
  \begin{subfigure}{0.32\linewidth}
    \centering
    \includegraphics[width=\linewidth]{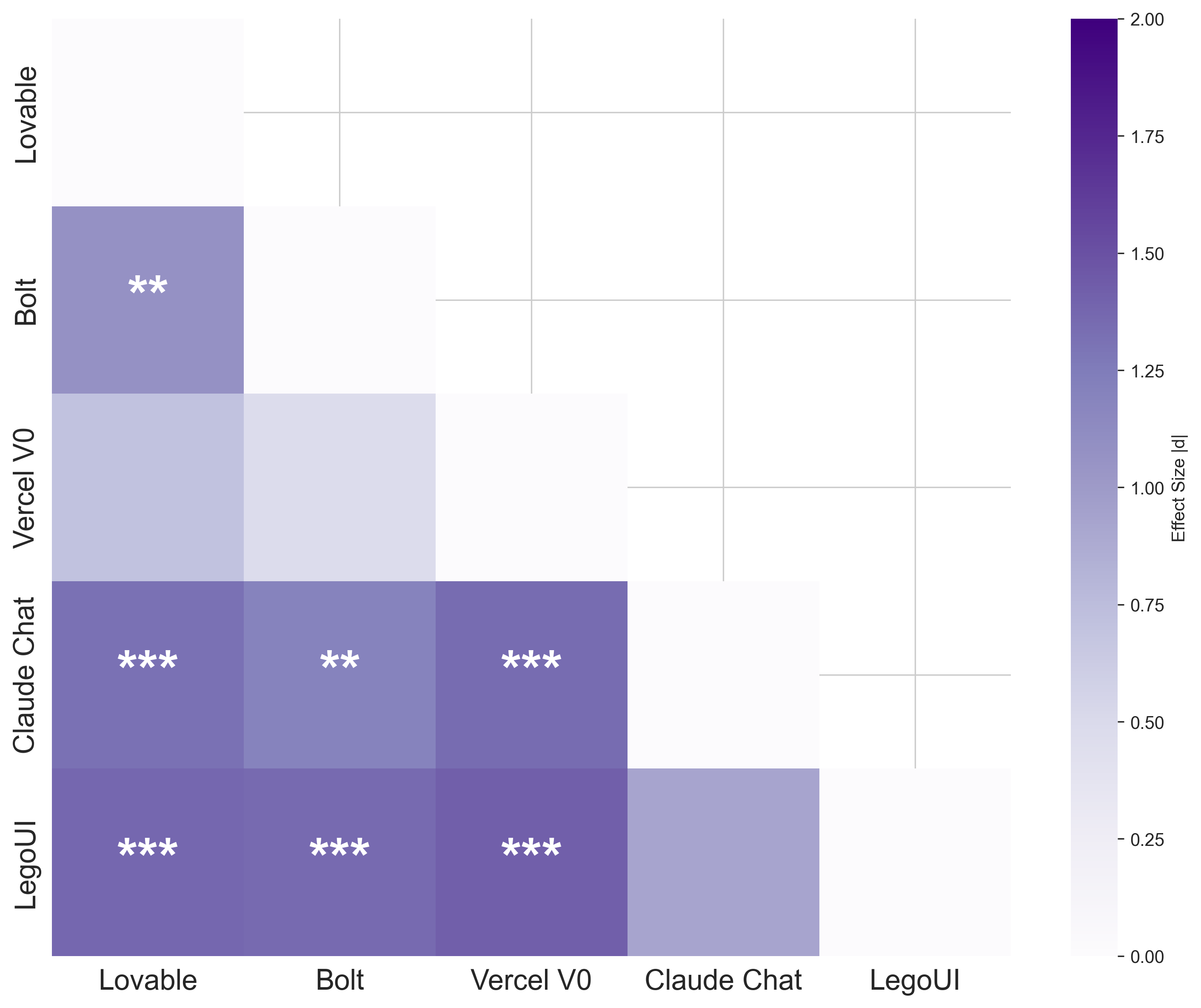}
    \caption{Visual Coherence}
    \label{fig:heatmap_visual}
  \end{subfigure}
  \hfill
  \begin{subfigure}{0.32\linewidth}
    \centering
    \includegraphics[width=\linewidth]{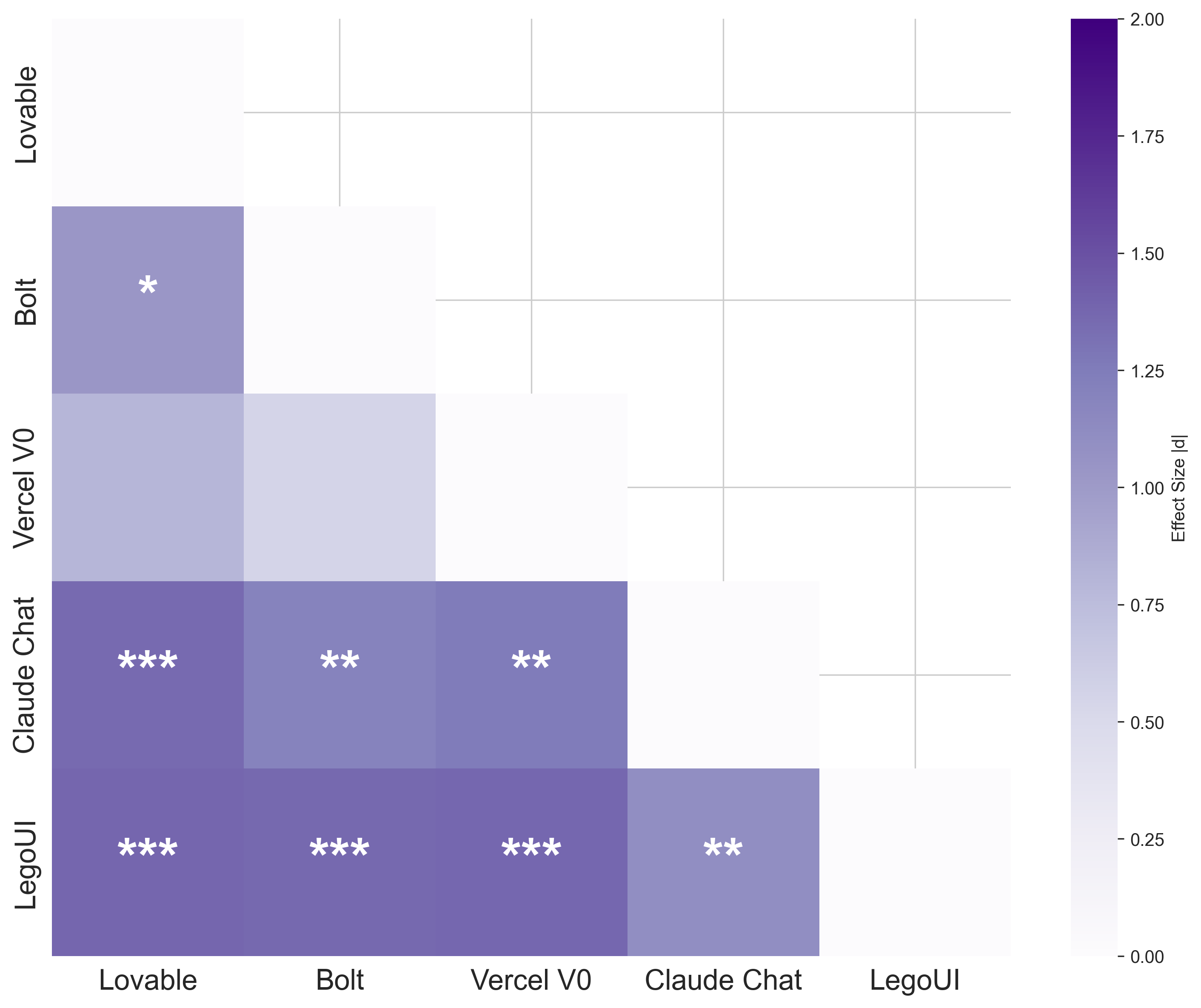}
    \caption{Trustworthiness}
    \label{fig:heatmap_trust}
  \end{subfigure}

    \caption{Comparative results of UI evaluations across five dimensions. 
    (a) shows the overall average score matrix by dimension, where each cell reports the mean score of a system on a given dimension and highlights cross-system comparisons and relative strengths. 
    (b–f) present pairwise $t$-test heatmaps for each evaluation dimension, with shading indicating effect size and significance levels.}

  \label{fig:ui_results}
\end{figure}

Participants rated the anonymized UIs on five quality dimensions: intent alignment, usability, accessibility, visual coherence, and trustworthiness. Figure~\ref{fig:UI_Dimensions_Comparison} presents the mean scores with standard deviations across these dimensions. To complement this view, Figure~\ref{fig:ui_avg_scores} shows the full score matrix with cross-system rankings. Full results of the evaluation from each system are provided in Table~\ref{tab:UI_scores}.

Across all dimensions, the UIs generated by \tool{} received the highest ratings, with averages exceeding 4.2 on every measure. Its strongest dimensions were visual coherence ($M=4.53$) and accessibility ($M=4.49$). By contrast, \textit{Lovable} was consistently rated lowest, with scores ranging from 2.7 to 3.2 across dimensions. While it sometimes achieved relatively strong scores on specific prompts (e.g., intent alignment $M=4.13$ in the portfolio scenario), its average remained lower since competing systems consistently reached higher levels across dimensions.

Among the industry baselines, \textit{Bolt} emerged as the strongest overall compared to \textit{Lovable} and \textit{Vercel V0}. Its mid-range performance ($M=3.60$ for accessibility, $M=3.47$ for visual coherence, $M=3.29$ for trustworthiness) ranked above \textit{Lovable} and slightly higher than \textit{Vercel V0}. While \textit{Vercel V0} occasionally outperformed Bolt on individual prompts, for example showing higher usability in the productivity task ($M=3.67$ vs.\ $M=3.33$), its overall averages were slightly lower, especially on trustworthiness ($M=3.09$).

Rather than relying only on mean differences, we also examined pairwise significance tests, summarized in Figure~\ref{fig:heatmap_intent}--\ref{fig:heatmap_trust}
. The heatmap shows that \textit{\tool{}} was rated higher than the other systems on all five dimensions. \textit{Claude Chat} was also reliably above the baseline one-shot tools in several measures, placing it between the \tool{} and the industry tools. Within industry tools, differences were smaller, though \textit{Bolt} showed a steadier edge, particularly compared to \textit{Lovable}.

\begin{figure*}[ht]
  \centering

  \begin{subfigure}{0.48\linewidth}
    \centering
    \includegraphics[width=\linewidth]{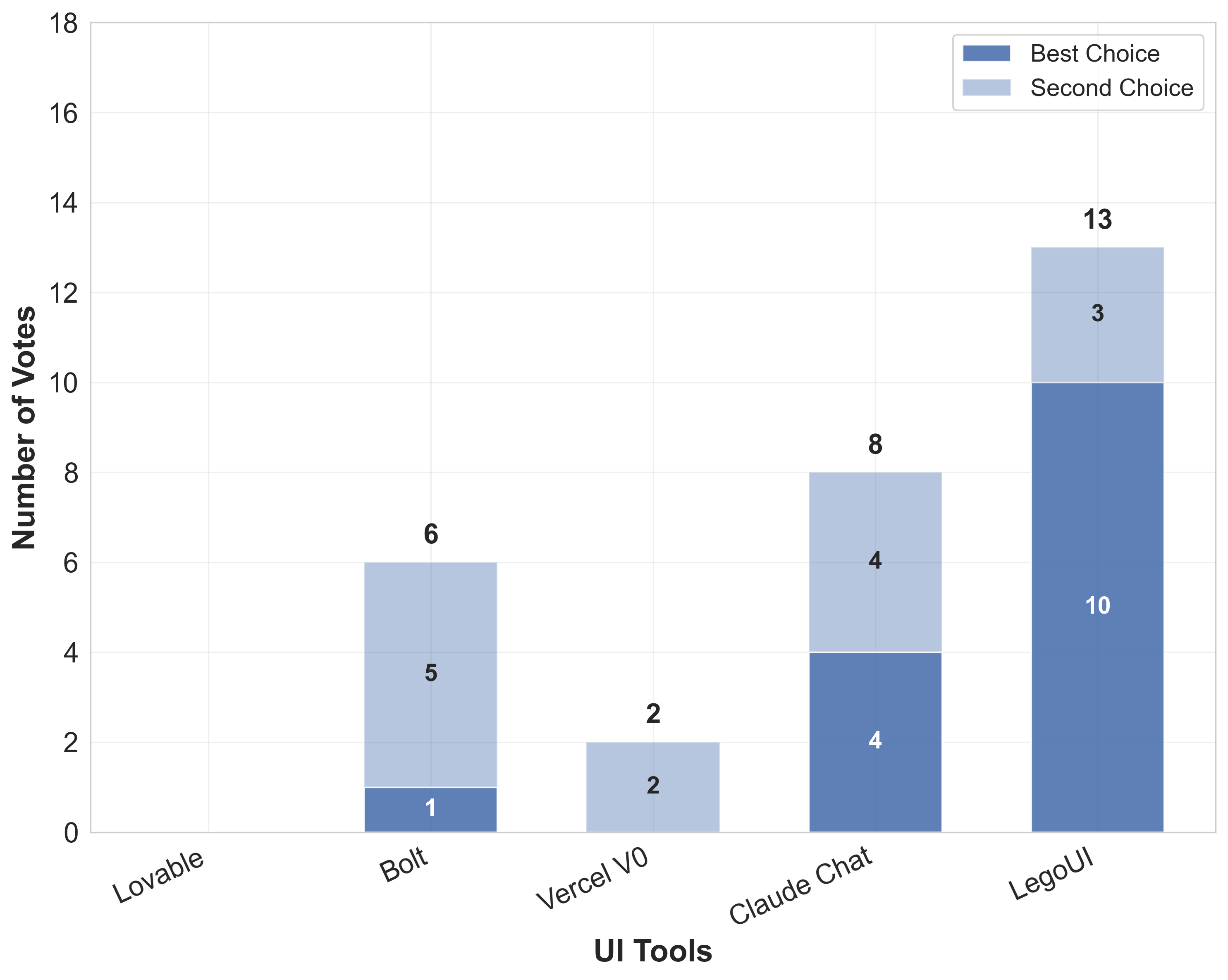}
    \caption{Dashboard prompt preferences}
    \label{fig:dashboard_pref}
  \end{subfigure}
  \hfill
  \begin{subfigure}{0.48\linewidth}
    \centering
    \includegraphics[width=\linewidth]{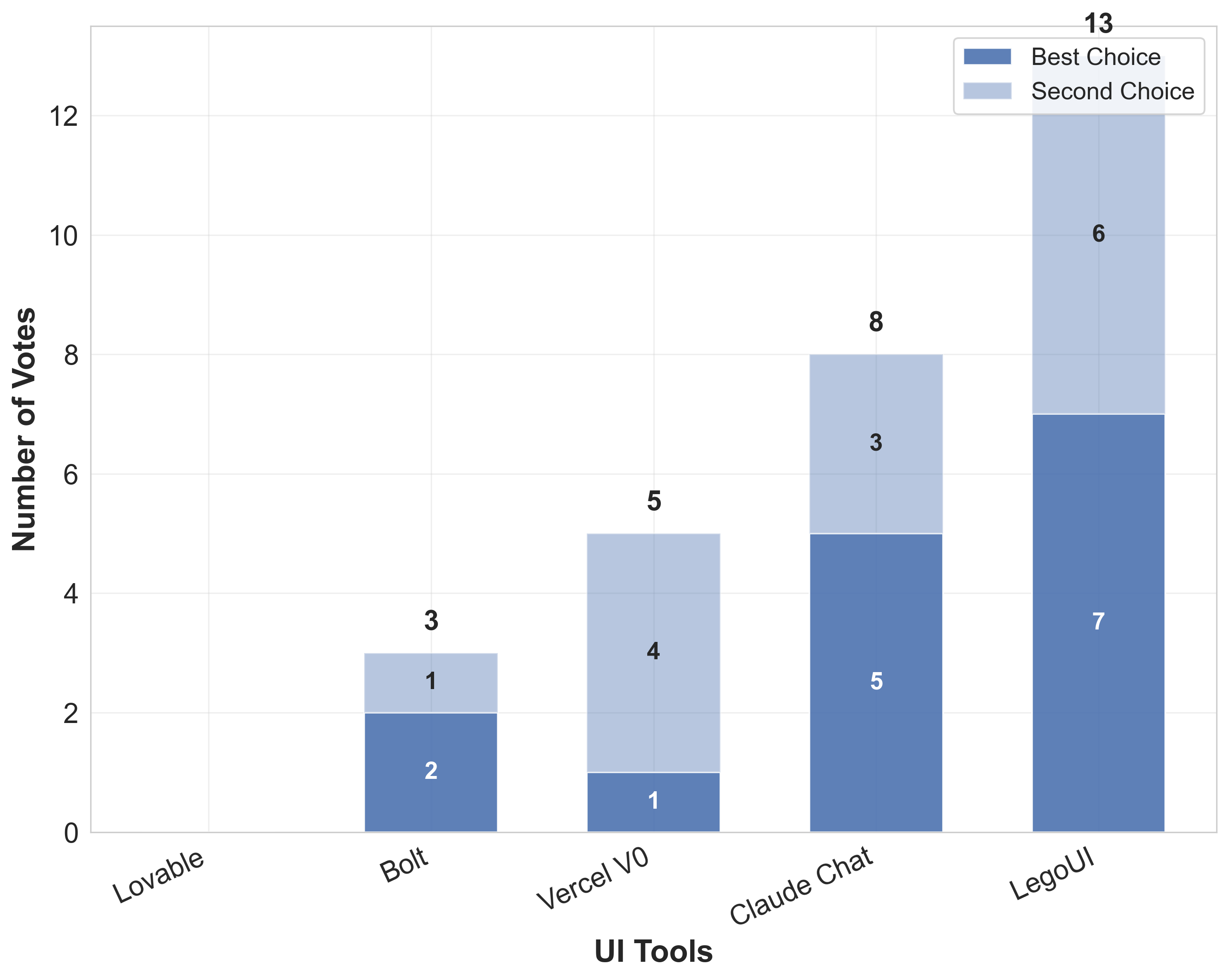}
    \caption{Productivity prompt preferences}
    \label{fig:productivity_pref}
  \end{subfigure}

  \begin{subfigure}{0.48\linewidth}
    \centering
    \includegraphics[width=\linewidth]{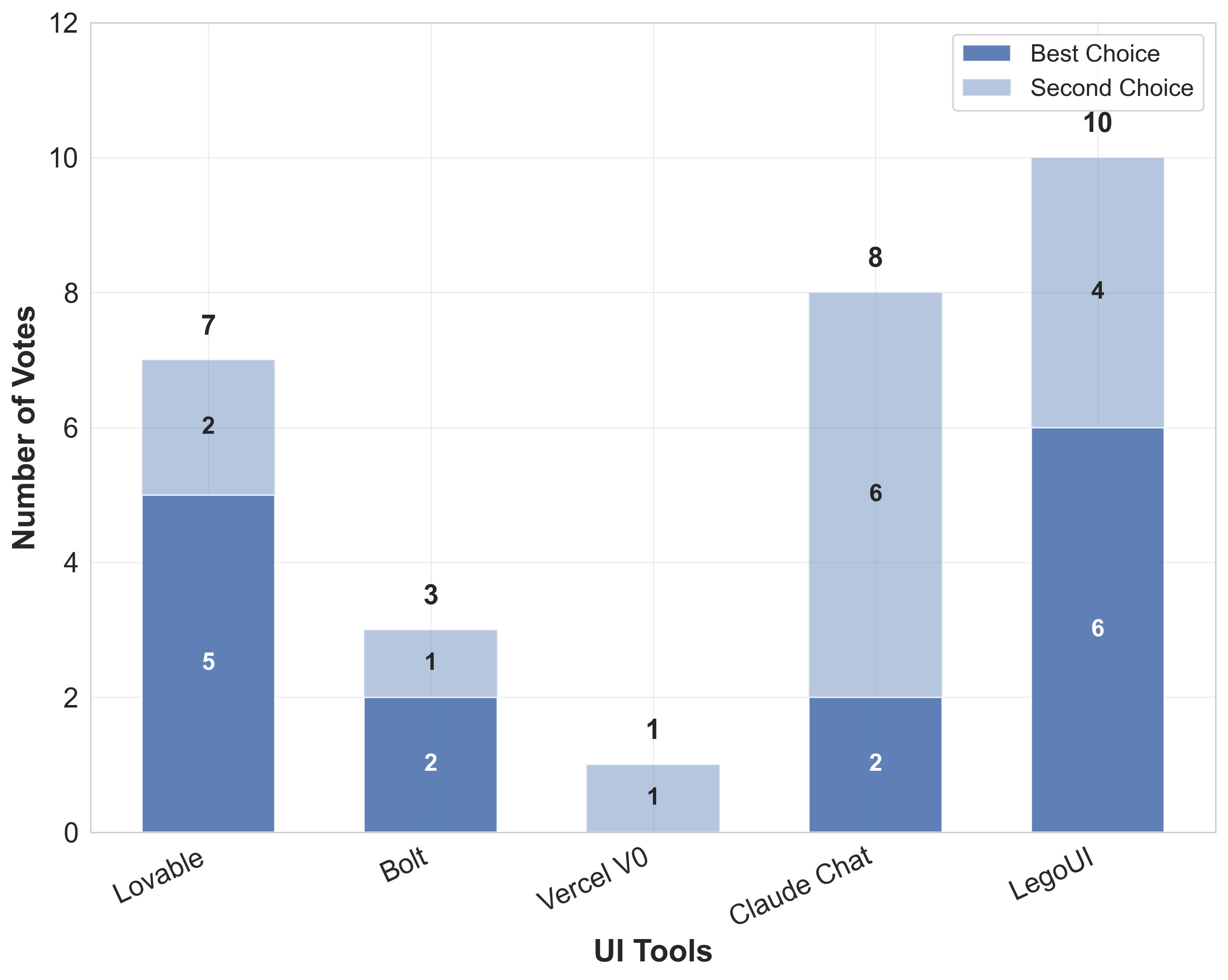}
    \caption{Portfolio prompt preferences}
    \label{fig:portfolio_pref}
  \end{subfigure}
  \hfill
  \begin{subfigure}{0.48\linewidth}
    \centering
    \includegraphics[width=\linewidth]{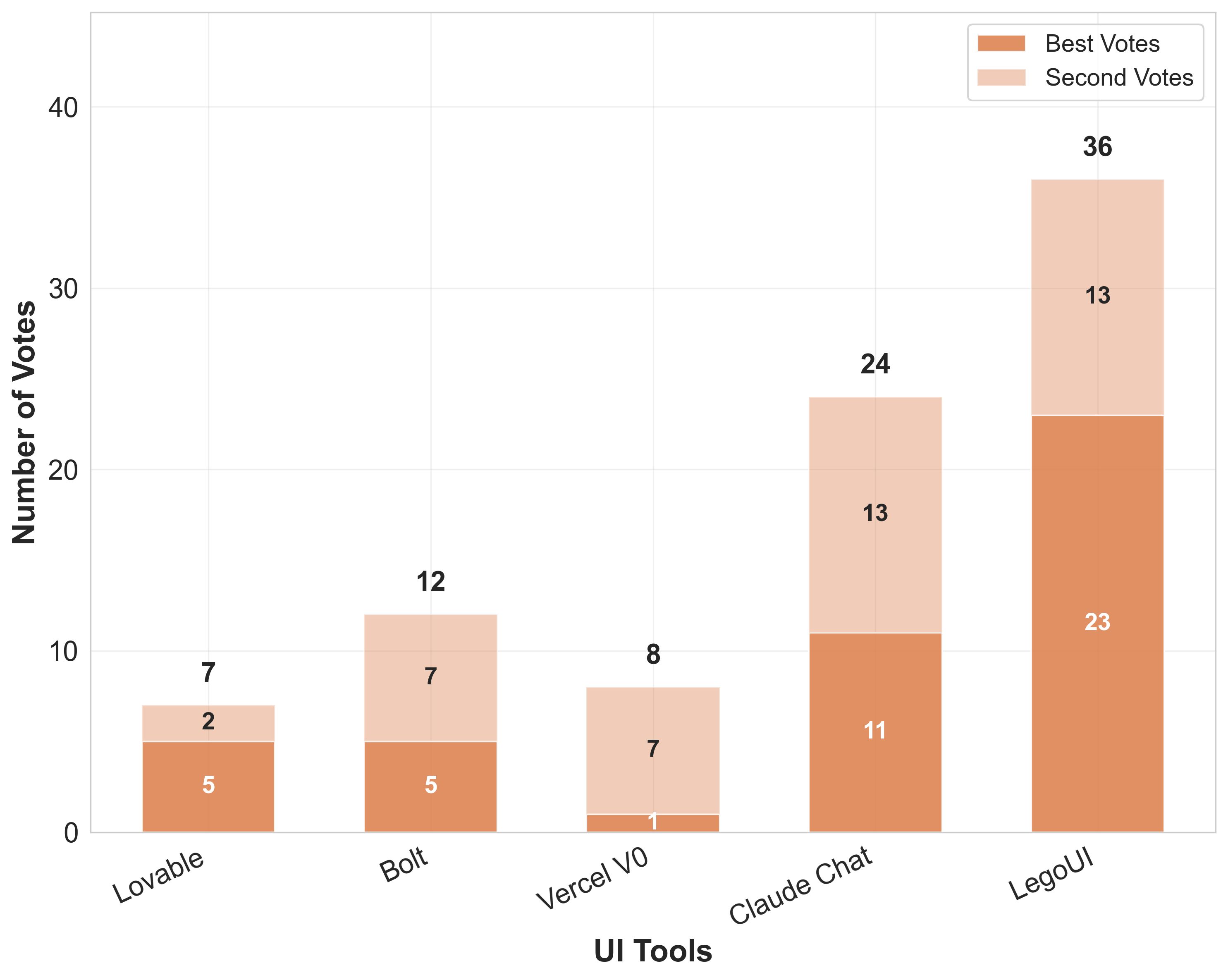}
    \caption{Overall preference rankings across all prompts}
    \label{fig:final_rankings}
  \end{subfigure}

  \caption{User preference results across prompts. (a–c) show best-choice and second-choice votes for Dashboard, Productivity, and Portfolio prompts respectively. (d) summarizes overall rankings aggregated across all prompts.}
  \label{fig:ui_pref_results}
\end{figure*}

In addition to numerical ratings, we also asked participants to select their first and second preferred UIs for each prompt. 
Figures~\ref{fig:ui_pref_results}(a–c) showed the distribution of preference votes for Dashboard, Productivity, and Portfolio prompts respectively, while Figure~\ref{fig:ui_pref_results}(d) summarizes the overall rankings across all prompts. \textit{\tool{}} received the highest number of first-choice votes (23) and additional second-choice votes (13) in overall preference ranking. \textit{Claude Chat} followed with 11 first-choice and 13 second-choice votes. Among the industry tools, \textit{Bolt} attracted the highest number of preferences (5 first-choice, 7 second-choice), while \textit{Vercel V0} and \textit{Lovable} were selected less often overall.

\subsubsection{Task 2 Results: \tool{} Interaction Evaluation}
\begin{figure}
    \centering
    \includegraphics[width=1\linewidth]{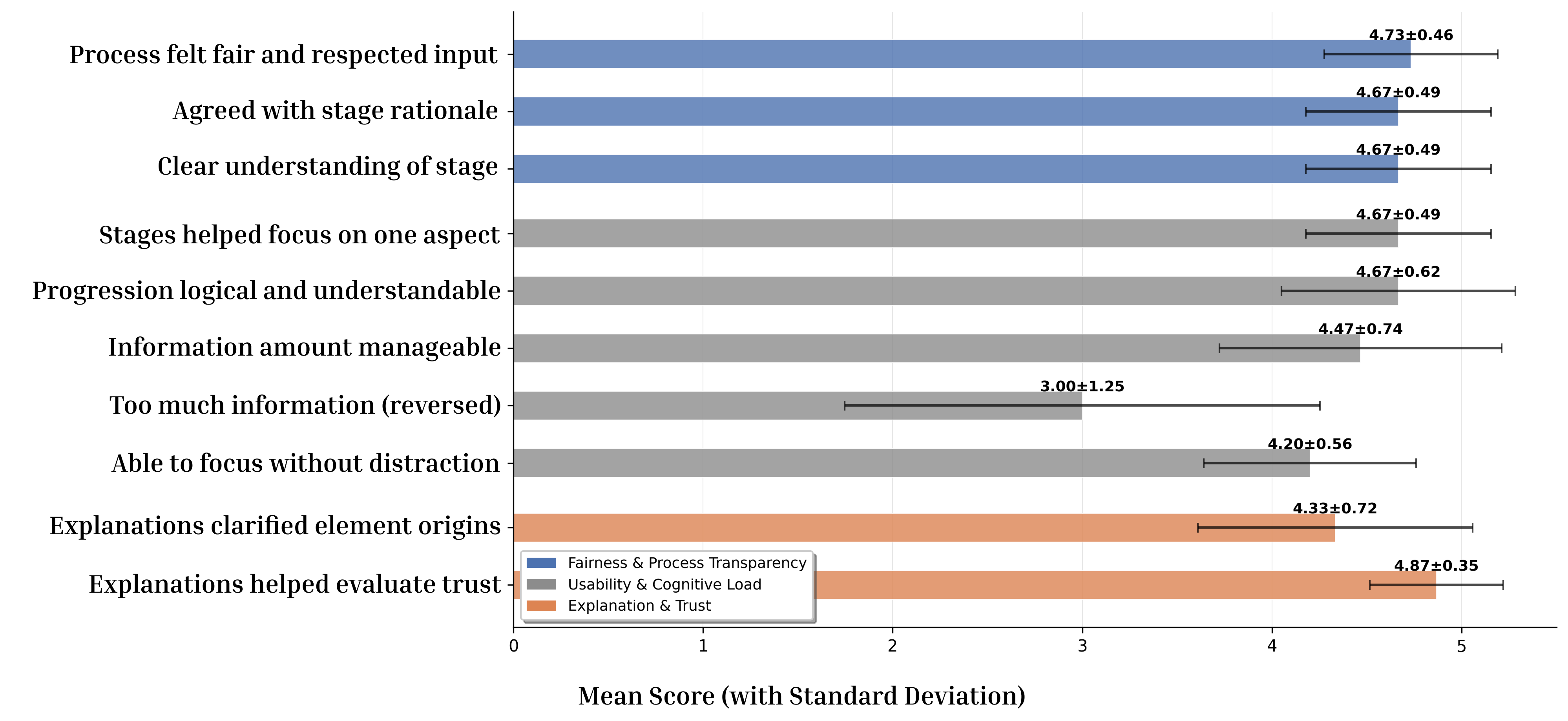}
    \caption{Mean scores with standard deviations for all questionnaire items, grouped by dimension (Fairness \& Process Transparency, Usability \& Cognitive Load, Explanation \& Trust).}
    \label{fig:interaction-questionnaire}
\end{figure}
To evaluate participants’ experience of interacting with the \tool{}, we administered a questionnaire covering three dimensions: 1) fairness and process transparency, 2) usability and cognitive load, and 3) explanation and trust. Mean ratings across all items were high, with responses concentrated in the upper range of the scale (\autoref{fig:interaction-questionnaire}; full distributions are included in Appendix~\ref{A: interaction}). 

\paragraph{Fairness and Transparency:} Most participants agreed that the system respected their input and that the rationale of each stage was clear (11 chose “strongly agree,” 4 chose “agree”). Participants also indicated that they could easily track which stage they were in during the workflow (RQ2).  

\paragraph{Usability and Cognitive Load:} The staged process was seen as helpful for focusing on one aspect of the UI at a time. The majority agreed that the progression from one stage to the next was logical and that the amount of information shown was generally manageable. Responses to the reverse-coded item on information overload were more varied. More than half of the participants rated it low, but several gave scores of 4 or 5. This indicates that some users felt the system occasionally presented more information than they preferred (RQ2).

\paragraph{Explanation and Trust:} Most participants agreed that the system’s explanations clarified how elements were derived and helped them assess the output (RQ1, RQ2).

\subsubsection{Task 3 Results: Post-Generation Editing}
We further assessed participants’ experience of post-generation editing through a set of questions covering control, explanation and transparency, intent alignment, usability and trust, and task effectiveness and exploration (Figure~\ref{fig:five_dimensions_comparison}; full distributions in Appendix~B).  

Overall, \textit{\tool{}} received the most favorable evaluations. Participants frequently rated it highly on items concerning meaningful control, the ability to adjust undesired decisions, and the system’s responsiveness to feedback. This indicates that the staged workflow from \tool{} afforded users more agency in editing (RQ3). High scores on explanation-related items (e.g., understanding why each UI element appeared, distinguishing prompt-specified versus inferred content) further suggest that the \tool{}'s scaffolded design reduced ambiguity during modification (RQ3).  

By contrast, responses for the one-shot systems reflected more constraints. \textit{Claude Chat} was rated around the midpoint on control-related questions (M=2.96), indicating limited ability to adjust or undo system choices. For \textit{Bolt}, ratings were higher in task efficiency (M=3.60) compared with \textit{Claude Chat}: participants agreed that it could help complete a design quickly, but elements in intent alignment, such as whether the final UI preserved the intended priorities, received more mixed responses (RQ3). 
These findings are consistent with prior studies, which similarly reported that one-shot tools offered speed but constrained fine-grained control and intent preservation~\cite{chen2025genui}. In both cases, participants emphasized that efficiency was often achieved at the expense of iterative flexibility.

\begin{figure}
    \centering
    \includegraphics[width=1\linewidth]{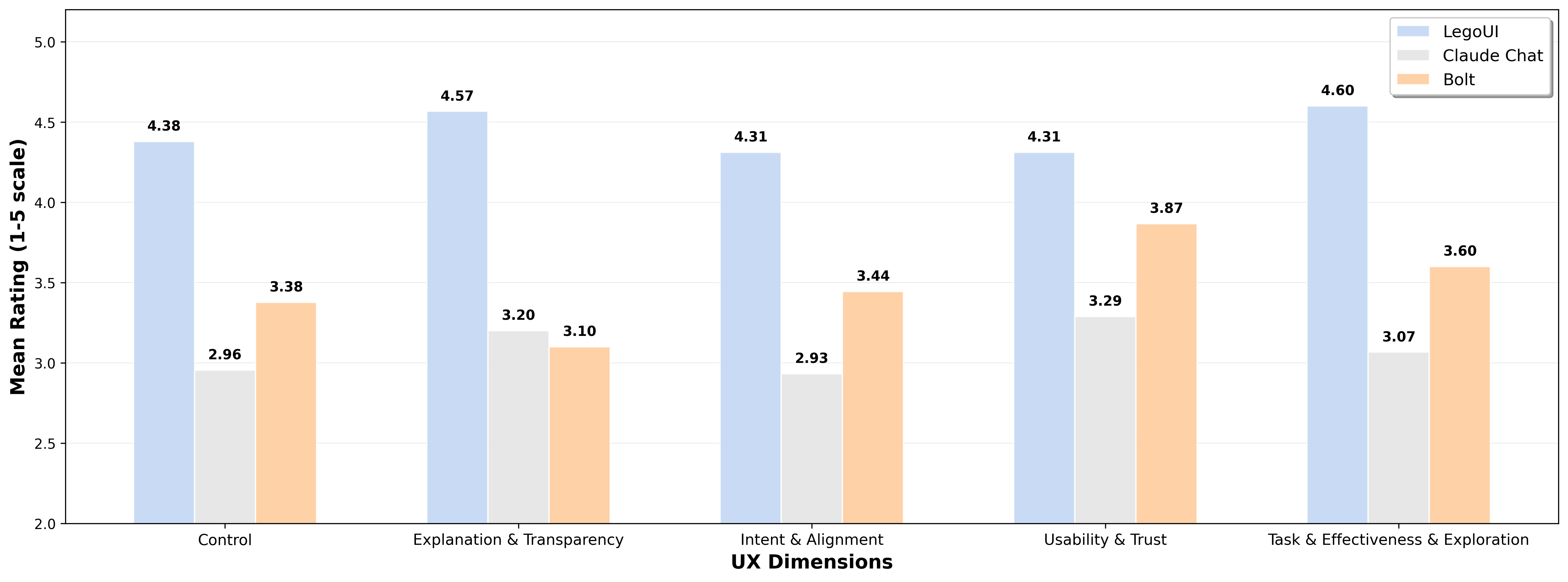}
    \caption{Mean ratings (1–5 scale) of post-generation editing experience across five UX dimensions: control, explanation and transparency, intent alignment, usability and trust, and task effectiveness and exploration.
}
    \label{fig:five_dimensions_comparison}
\end{figure}

\subsubsection{Interview}\label{subsubsec:interview}

From the interviews, the findings revealed several consistent patterns that corroborate the results from the questionnaires from task 2 and task 3.

First, participants (P1-3, P5, P10–P14) repeatedly emphasized the sense of control and agency they experienced with \tool{}. Compared to baseline one-shot system, Claude Chat and Bolt, they felt that \textit{\tool{}} allowed them to guide the process more actively and maintain oversight over how their inputs shaped the outcome. Decomposing the design process into requirements, layout, interaction, relation, and style was described as giving designers leverage over both detailed and higher-level design choices. As P3 reflected:
\ul{\textit{``
It helps me control what I need at every stage of the process instead of just generating the result for me.''}}

Second, our \tool{}, with staged workflow was associated with a strong perception of transparency and fairness. As P7 commented: \ul{\textit{``
The staged system supported my process better because its step-by-step flow kept decisions clear and consistent.''}} Participants (P3, P9–P12) also highlighted the rationale of each step was more visible than in one-shot systems, which made them feel that their input was taken seriously and consistently respected. 

Third, several participants (P1, P2, P7-10, P14) noted that the clarity of stage-by-stage progression supported their editing strategies. As P14 expressed
\ul{\textit{``This system provides a step-by-step and fine-grained control. For each stage, it guides me to focus on one aspect specifically. (P14)''}} 

However, two participants (P12, P15) show their preference on Bolt. They mentioned the expression of Bolt is easy to understand and the results generate by Bolt is the most accurate and complete.

\subsection{Findings}

Our evaluation across three tasks and follow-up interviews highlights how participants used \tool{} differently from one-shot GenUI systems. Rather than only comparing output scores, we focus on how staged generation changed participants' understanding, judgment, and intervention during UI creation.

\subsubsection{Quality as Alignment Rather Than Visual Polish}

\tool{} received strong ratings and preference votes across the evaluated prompts. Participants rated it highest on visual coherence, accessibility, and intent alignment, and most frequently selected it as their preferred output. These results indicate that participants valued the generated interfaces as coherent responses to the design goals, rather than judging quality only by visual richness.

The comparison with industry tools clarifies this pattern. Bolt and Claude Chat sometimes produced more polished first-pass outputs, and some participants appreciated Bolt for being easy to understand, accurate, and complete. These preferences show that visual polish and implementation readiness mattered during quick evaluation. At the same time, \tool{} was rated more consistently across prompts and dimensions. This suggests that staged specification helped stabilize the relation between the prompt and the generated interface. Participants appeared to value interfaces that preserved the intended structure and purpose, even when other tools offered more immediately polished results.

This finding is also reflected in participants' comments about visual appeal. For example, P15 described \tool{} as ``the most visually appealing,'' despite the system's simpler HTML-based output. This suggests that perceived quality came from coherence, clarity, and alignment with intent, rather than from richer visual effects alone.

\subsubsection{Transparency as Checking System Interpretation}

Participants used the staged workflow as a way to check how the system understood their prompt. The intermediate stages made generation more understandable because participants could see how their input was translated into design decisions before the final UI appeared. This changed transparency from a general sense of system explainability into a more concrete activity of checking interpretation.

This was especially useful for beginner users who may not know how to fully specify a UI upfront. Instead of needing to write a perfect prompt, they could review the system's interpretation step by step. The staged process gave them moments to confirm whether the extracted requirements, components, layout decisions, interactions, relations, and styles matched what they intended. P14 described this as ``step-by-step and fine-grained control,'' adding that each stage guided them to ``focus on one aspect specifically.'' This shows that transparency helped participants organize their attention and make sense of the design process.

Participants also associated this visibility with fairness. They could see which decisions came from their own input and which were inferred by the system. This distinction mattered because inferred decisions could be helpful, yet they could also move the design in a direction the user had not intended. One-shot systems were often appreciated for speed and completeness, while \tool{} gave participants more opportunities to inspect the path from prompt to output.

\subsubsection{Control as Early Intervention}

Participants valued \tool{} because it allowed them to intervene before design decisions were embedded in the final UI. The accept, reject, and add operations gave participants a way to shape the design state during generation. This changed editing from a mainly corrective activity into an earlier process of guiding assumptions.

In one-shot systems, participants typically reacted to a completed interface. Their work involved identifying what looked wrong, writing another prompt, and checking whether the new output had changed in the desired way. In \tool{}, participants could act on intermediate items before rendering. Accepting an item preserved a decision. Rejecting or revising an item prevented that assumption from carrying forward. Adding a new item allowed participants to introduce a requirement at the moment it became relevant.

The P4 workflow illustrates this pattern. During staged generation, P4 rejected several AI-generated specifications, added a custom relation, and added a styling directive for a minimalist layout. Later, when generated ``Continue reading'' links pointed to the same placeholder page, P4 revised the specification by clarifying that each link should lead to a distinct content page. This case shows how participants used staged control to diagnose and correct assumptions in the design state, rather than only describing surface-level changes to the final artifact.

\subsubsection{Different Stages Supported Different User Judgments}

Participants did not experience all stages as equally easy to judge. The staged workflow helped them focus on one aspect of the design at a time, yet the usefulness of each stage depended on whether participants could understand the decision being shown and anticipate its effect.

Component-level decisions were often easier to evaluate because they were close to the task goal. Participants could usually judge whether a UI should include a button, card, filter, image, or navigation element. Layout decisions became easier when participants could connect them to the emerging visual structure. These stages supported recognition-based judgment, where users could decide whether a proposed decision matched what they wanted.

Interaction and relation stages required more effort. Their effects were distributed across components, pages, or states, making them harder to evaluate from text alone. Participants sometimes needed additional context to understand what an interaction or relation would change in the interface. This suggests that staged generation helped organize design judgment, while also revealing that some design dimensions require more visual or behavioral grounding.

This finding is important because it shows that user control depends on the form of the decision. Participants were more confident when decisions were concrete and locally judgeable. They needed more support when decisions were abstract, cross-component, or dependent on later behavior.

\subsubsection{Staged Generation Shifted the Cost of Design Work}

Participants generally found the staged process logical and manageable, and many appreciated the explicit references, source highlighting, and visible links between UI elements and specifications. These features helped them understand where information came from and how decisions were derived.

At the same time, several participants described moments of information overload. The staged workflow required them to review more intermediate details than one-shot systems. This added inspection work, especially when inferred decisions accumulated across stages or when the meaning of an item was abstract.

This trade-off suggests that \tool{} did not simply reduce work. It shifted part of the work earlier in the process. One-shot systems lowered the effort needed to start and supported faster interaction, yet they placed more burden on prompt accuracy and post-generation correction. \tool{} required more inspection during generation, yet this inspection allowed participants to catch and adjust assumptions before they shaped the final output. For beginner users, this shift can be valuable when the system provides enough structure to support judgment. It can become costly when users are asked to inspect details without enough context about their importance or consequence.

\section{Discussion}

Our findings suggest that the value of staged GenUI lies in changing how users participate in generation. In many prompt-to-UI tools, prompt interpretation, model inference, and artifact generation are tightly coupled. Users can revise the final output, while the assumptions behind that output remain difficult to inspect or adjust. \tool{} separates these assumptions into staged UI-DSL items, making them available for review and revision before they are rendered into an interface.

We discuss how this staged structure changes the user's role in GenUI. We first frame intermediate representations as negotiable design commitments. We then discuss useful forms of transparency, the relationship between control granularity and user judgment, the position of staged control within the design process, the cost of inspectability, and implementation directions for future systems.

\subsection{From One-shot Generation to Negotiating Design Commitments}

A key implication of our study is that intermediate representations become useful when they give users a place to act during generation. In \tool{}, UI-DSL items make the system's emerging design assumptions visible as provisional commitments. These commitments may describe a component, layout structure, interaction rule, relation, or style direction. Users can accept, reject, or revise them before they become part of the final interface.

This changed how participants engaged with the system. In one-shot workflows, users typically evaluate a completed artifact and then describe what should change. In \tool{}, participants could intervene while the design state was still being constructed. Accepting an item preserved a decision for later stages. Rejecting or revising an item changed the assumptions available to subsequent generation. Control came from shaping the conditions under which the interface would be generated, instead of only correcting the generated interface afterward.

This process can be understood as negotiation around design commitments. The system proposes possible interpretations and extensions of the user's intent. Users then decide which proposals should remain part of the active design state. This negotiation matters because many UI decisions are under-specified in natural language prompts. Users may describe a desired experience, content, or style while leaving layout structure, relations, and interaction behavior open to interpretation. Staged representations made these interpretations available for review before they became embedded in the output.

This framing clarifies the role of the UI-DSL. It provides a shared design state where user intent, model inference, and user revision can accumulate across stages. For GenUI systems, this suggests a form of controllability that sits between prompt rewriting and final-output editing. Staged design commitments allow users to inspect and revise the assumptions that connect intent to output.

\subsection{Forms of Transparency in Prompt-to-UI Generation}

Our findings suggest that transparency in prompt-to-UI generation is most useful when it helps users evaluate the system's interpretation of their intent. Participants needed to see how their prompt was translated into design decisions, where the system filled in missing details, and how those decisions would shape later generation.

We identify three forms of transparency from this process. \textit{Interpretation transparency} concerns how the system decomposes a prompt into design requirements. In \tool{}, the requirement analysis stage showed how user input was represented across design dimensions. This allowed participants to check whether the system had captured the main intent before moving forward. \textit{Inference transparency} concerns what the system adds beyond explicit user input. Prompt-to-UI generation often requires the model to infer underspecified details, and making these inferences visible allowed users to decide whether they aligned with the intended design direction. \textit{Consequence transparency} concerns how an intermediate decision affects later parts of the interface or generation process.

The current system supported interpretation and inference transparency more clearly than consequence transparency. Participants valued seeing what decision was made at each stage, yet some struggled when the effect of a decision was distributed across components, pages, or behaviors. This was especially visible in relation and interaction stages. A navigation decision may depend on whether a destination page exists. A relation item may influence several components at once. In these cases, item text gave users limited support for judging what the decision would do.

Transparency should help users move from inspection to action. Provenance can show whether an item came from the prompt, model inference, or user edit. Item state can show whether it remains active, rejected, or revised. Consequence should be made visible by linking a decision to affected UI regions, components, behaviors, or later stages. With this connection, users can evaluate a decision through its role in the evolving interface, rather than through textual description alone.

For staged GenUI, transparency should be selective and consequence-oriented. Showing every intermediate item with equal emphasis may increase visibility while adding little support for user judgment. A more useful approach is to expose information that helps users decide whether a design commitment should remain active, especially when its effects are difficult to infer from text alone.

\subsection{Control Granularity and Users' Design Judgment}

Our results suggest that staged control is effective when users can make a meaningful judgment about a design decision at the moment it is presented. \tool{} separates generation into design dimensions, yet participants experienced these stages as points where they could evaluate, confirm, or redirect the system's interpretation of their intent. This changed their role from judging a completed AI-generated artifact to participating in the formation of the design state.

This shift was most useful when the staged item matched a judgment users could confidently make. Component-level decisions often worked in this way. Users could decide whether an interface should include a search bar, card, filter, button, or navigation element because these decisions were close to their task goals. In these cases, AI assistance helped by externalizing possible requirements and giving users a structured opportunity to recognize and refine design commitments.

Layout decisions involved a different kind of judgment. Participants could often evaluate a proposed layout once it became visible, even when they had described it only vaguely in the initial prompt. This suggests that AI can help users move from underspecified spatial intent to concrete design judgment. For layout, staged control becomes useful when it supports recognition, comparison, and visual grounding.

Interaction and relation decisions revealed the boundary of staged control. These decisions required participants to reason about consequences that were harder to perceive immediately. A user may know that a button should exist, then struggle to judge whether a relation between pages, states, or components is appropriate without seeing its behavioral effect. In these cases, the task shifts from design selection to abstract reasoning, increasing cognitive effort and weakening the sense of control.

This distinction helps explain how AI changes user behavior in GenUI workflows. In one-shot tools, users often respond after the model has already made many design decisions. Their work becomes corrective, involving noticing what looks wrong, revising the prompt, and generating again. In \tool{}, users could intervene earlier. Their work became more diagnostic, involving examination of system assumptions, selection of aligned assumptions, and prevention of unsuitable assumptions from shaping later output. This earlier intervention was valuable when users could understand the decision they were asked to judge.

GenUI systems should expose control when users are positioned to act on it. Some decisions are locally judgeable through the item itself or a nearby preview. Other decisions depend on components, pages, behaviors, or visual outcomes. For dependency-based decisions, the system should support users by showing consequences, examples, or affected areas instead of presenting the decision as an isolated checklist item.

This reframes staged generation as a user-centered scaffolding problem. Stages slow down generation at points where human judgment can improve the design. They provide less value when users are asked to approve abstract system logic without enough perceptual or contextual support. Future staged GenUI systems should adapt the form of control to the kind of judgment users can make, helping users participate where their judgment is strongest and providing additional support where AI-generated decisions are difficult to evaluate directly.

\subsection{Positioning Staged Control Within the Design Process}

Our results suggest that staged control and one-shot generation support different user needs across the design process. Early in a task, users may still be exploring possible directions. An immediate generated artifact can help them see possibilities, compare styles, and clarify preferences. As expectations become clearer, users need stronger support for preserving intent, revising specific decisions, and maintaining consistency across iterations.

This helps explain why participants valued different tools for different purposes. One-shot tools supported quick first passes, polished visual examples, and implementation-oriented output. They lowered the effort needed to start. Staged control became more valuable when users wanted to understand how the interface was being constructed, prevent unwanted assumptions from carrying forward, or refine the result through multiple rounds. At this point, the design task shifted from producing an artifact to managing the assumptions behind it.

This distinction also shows how AI changes the user's role across the workflow. During exploration, users may benefit from letting the model make more decisions because the goal is to generate possibilities. During refinement, users need model decisions to become inspectable and revisable because the goal is to align system assumptions with more specific design intent. Staged control is most useful when users move from open-ended exploration toward commitment and refinement.

Future GenUI systems could support this transition more explicitly. A \textit{preview-first} workflow would let users generate an initial interface quickly and then enter staged refinement after identifying what should change. A \textit{stage-first} workflow would let users inspect and confirm requirements before rendering, which is useful when intent preservation matters from the beginning. GenUI systems could allow users to move between these workflows as their design intent becomes clearer.

Effective GenUI tools should support movement between exploration and commitment. The appropriate level of control depends on where users are in the design process, how clearly they can articulate their intent, and how much they need later outputs to preserve earlier decisions.

\subsection{Balancing Inspectability and Interaction Cost}

Making intermediate decisions visible also changes the work users need to do. Participants generally found the staged process logical and manageable, yet some described it as long or information-heavy. The cost of staged GenUI comes from repeated inspection and judgment. Users gain more opportunities to guide the system while also taking on the effort of deciding which intermediate decisions deserve attention.

The central question is whether visible information supports action. An intermediate item is useful when users can judge its relevance, anticipate its effect, and decide how to revise it. It becomes costly when users need to read many items without knowing their importance or consequence. This was most apparent in style, relation, and interaction stages, where textual items sometimes required users to infer effects that were easier to understand through visual or behavioral feedback.

Exposing every inferred decision with the same prominence may increase visibility while also increasing inspection work. Future systems should use selective and adaptive transparency by foregrounding decisions that users can meaningfully act on, linking items to affected UI regions or behaviors, and keeping less consequential details available in the background. The goal is to preserve the agency benefits of staged generation while reducing the effort required to understand and manage the design state.

\subsection{Implementation Limitations and Future Work}

The current implementation shaped how participants experienced staged control. \tool{} renders lightweight HTML mockups, which made generation stable and accessible for early-stage prototyping. Plain HTML also limited visual richness, data modularity, and integration with larger projects. Placeholder content reduced realism in some interaction tasks, such as when multiple ``Continue reading'' links pointed to the same mock page.

Industry tools often produced more polished first-pass outputs through style modules, animations, and component libraries. These outputs supported fast brainstorming and quick evaluation. They also sometimes introduced elements beyond the user's prompt, creating extra revision work. Our results suggest a trade-off between implementation fidelity and specification-level control. Higher fidelity is valuable when users want realistic prototypes, while staged specifications help preserve alignment with user intent.

Future versions of \tool{} could support modern frontend frameworks, modular data storage, and direct code editing while keeping the staged specification connected to the generated artifact. This connection preserves the design rationale behind higher-fidelity outputs. If a component, interaction, or style decision changes in code, the corresponding specification item should remain traceable and revisable.

Conflict handling is another direction. Late-stage edits may contradict earlier decisions, especially when users move across layout, interaction, relation, and style. The system could detect affected commitments, explain the conflict, and let users confirm a resolution. This would reduce manual tracking while preserving user control.

Finally, the staged sequence should become more flexible. Participants valued the separation of concerns, yet UI design decisions often depend on one another. Layout can shape interaction, interaction can require new components, and style can change perceived hierarchy. Future systems should preserve the clarity of stages while surfacing cross-stage dependencies when they become relevant.

\section{Conclusion}

As generative models are increasingly used in interface design, a central challenge lies in how design reasoning is represented and carried forward during generation. 
End-to-end approaches typically collapse generation into a single transformation, leaving the assumptions that guide design decisions implicit and difficult to revisit as design intent evolves.

This work presents a staged generation approach that models interface synthesis as an evolving reasoning process. By maintaining a structured specification throughout generation, design decisions become explicit artifacts that connect user intent with generated interfaces over time. The proposed UI-DSL serves as a persistent representation of this reasoning state, enabling inspection, revision, and continuation as generation progresses.

We instantiate this approach in LegoUI and evaluate it through a technical evaluation and a user study. The technical evaluation demonstrates that current large language models can derive structured specifications from natural language prompts. The user study shows that staged, specification-centered generation supports more transparent and manageable design interaction compared to one-shot baselines. These findings suggest that treating UI generation as an explicit reasoning process offers a viable foundation for supporting iterative design with generative models, while leaving open questions around how such representations can scale, adapt, and integrate with more complex design contexts.




\section*{Data Availability}

The technical evaluation data that support the findings of this study are available from the corresponding author upon reasonable request. The raw user study data are not publicly available due to privacy and ethical restrictions, as they contain information that could compromise participant confidentiality. De-identified user study data may be made available from the corresponding author upon reasonable request, subject to applicable ethical approval, institutional requirements, and participant confidentiality considerations.

\section*{Author Contributions}

Yinsi Zhou contributed to conceptualization, methodology, software development, investigation, data curation, formal analysis, validation, visualization, project administration, writing the original draft, and reviewing and editing the manuscript.

Mingyue Yuan and Hongyue Xu contributed to formal analysis, writing the original draft, and reviewing and editing the manuscript. Mingyue Yuan also contributed to conceptualization.

Jieshan Chen, Dong Wen, Xiwei Xu, Wenjie Zhang, Zhenchang Xing, and Gelareh Mohammadi contributed to methodology, formal analysis, supervision, and reviewing and editing the manuscript. Dong Wen, Xiwei Xu, Wenjie Zhang, Zhenchang Xing, and Gelareh Mohammadi also contributed to conceptualization. Zhenchang Xing and Gelareh Mohammadi also contributed to project administration.

Shidong Pan contributed to conceptualization, formal analysis, and reviewing and editing the manuscript.

Aaron Quigley contributed to visualization and reviewing and editing the manuscript.

All authors reviewed and approved the final version of the manuscript and agree to be accountable for all aspects of the work.

\printbibliography
\appendix
\newpage
\section{Illustrative Example of LegoUI}\label{A: example}
\begin{figure}[h]
    \centering
    \includegraphics[width=1\linewidth]{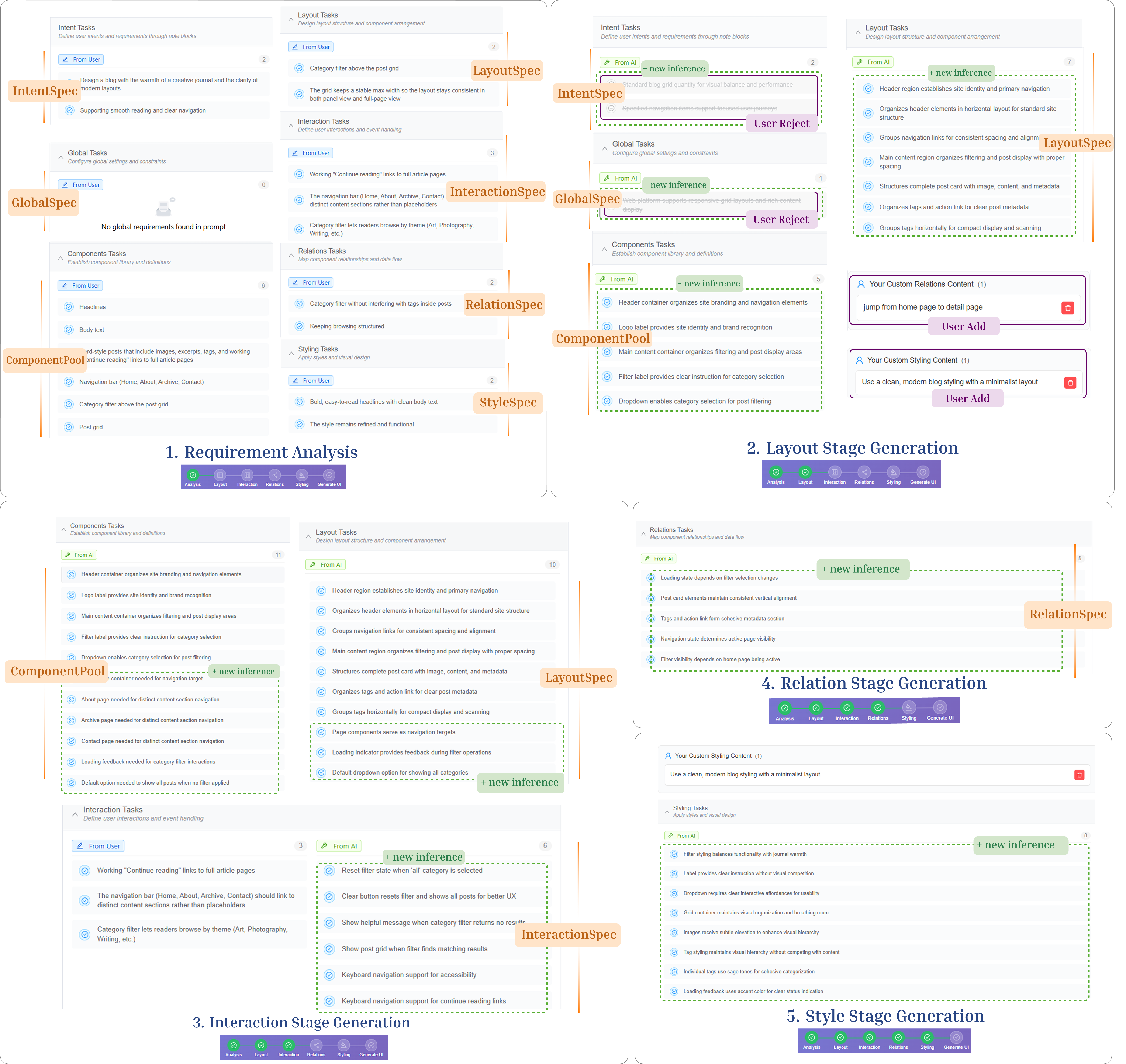}
    \caption{P4’s staged workflow across requirement analysis, layout, interaction, relation, and style stages.
The figure shows both AI inferences (green) and user interventions (purple), including rejections of intent and global specifications, as well as additions to relation and style specifications.}
    \label{fig:workflow-example}
\end{figure}

\begin{figure}
    \centering
    \includegraphics[width=1\linewidth]{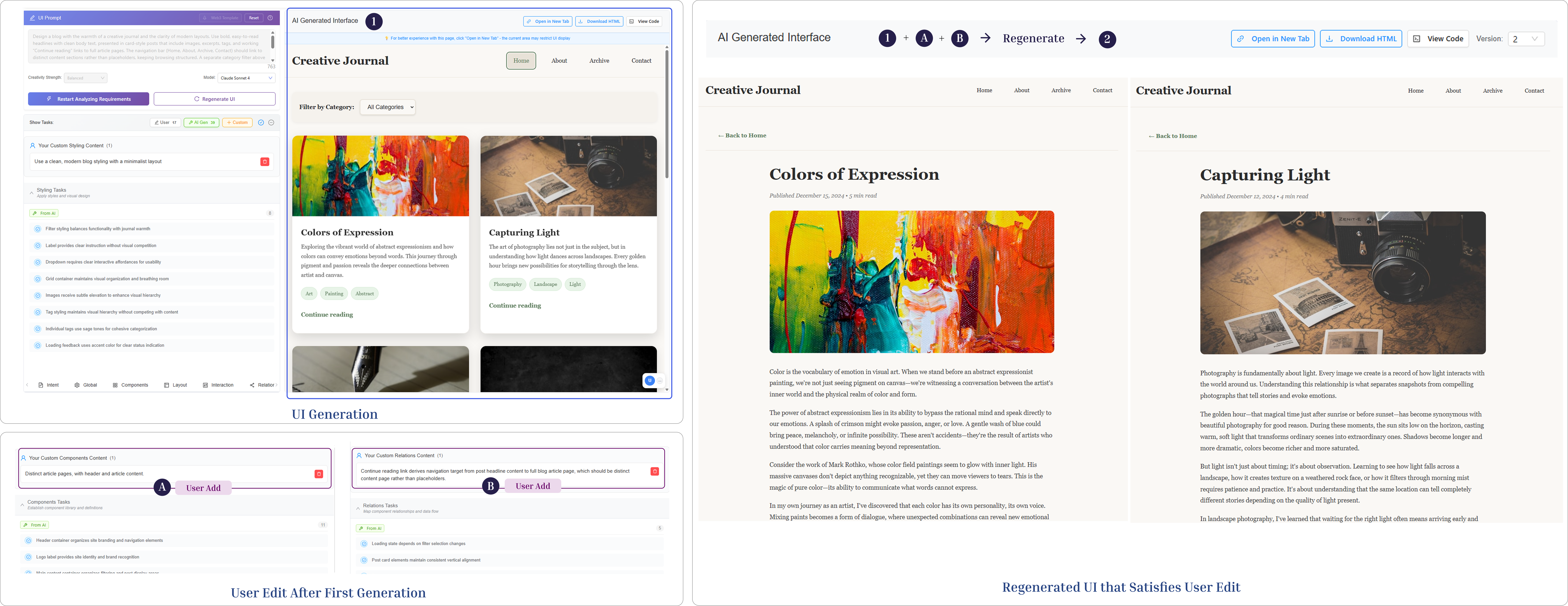}
    \caption{P4’s UI generation and subsequent refinement.
After the first UI draft was generated, the participant added new component and relation specifications (A, B) and then regenerated the interface, resulting in a revised UI that better matched their intentions.}
    \label{fig:UI example}
\end{figure}

To illustrate how participants engaged with intermediate results, we highlight the workflow of P4 (\autoref{fig:workflow-example}, \autoref{fig:UI example}). During the staged generation process, P4 actively intervened at several points. 
In the layout stage, P4 rejected both the AI-generated intent and global specifications, while adding a custom relation linking the home page to a global page and a custom styling directive for a minimalist layout (\autoref{fig:workflow-example}.1). 
In later stages, P4 accepted most AI inferences without further edits, but continued to review each specification. 

When the first UI version was generated (\autoref{fig:UI example}, left), P4 was dissatisfied with the lack of distinct article pages and navigation links. 
P4 therefore added new specifications: a component specifying \textit{“distinct article pages with header and article content”} (\autoref{fig:UI example}, A), and a relation specifying \textit{“continue reading links direct navigation from post headline to full article page”} (\autoref{fig:UI example}, B). 
After regenerating the interface, the system produced a second UI (\autoref{fig:UI example}, right) that integrated these edits, aligning more closely with P4’s intended design.

This case illustrates how P4 balanced acceptance of AI outputs with targeted edits across multiple stages, and how the final UI version emerged through iterative refinement.

\newpage
\section{LegoUI Interaction Task Questionnaire and Response Distribution} \label{A: interaction}
\begin{figure*}[h]
    \centering
    \includegraphics[width=1\linewidth]{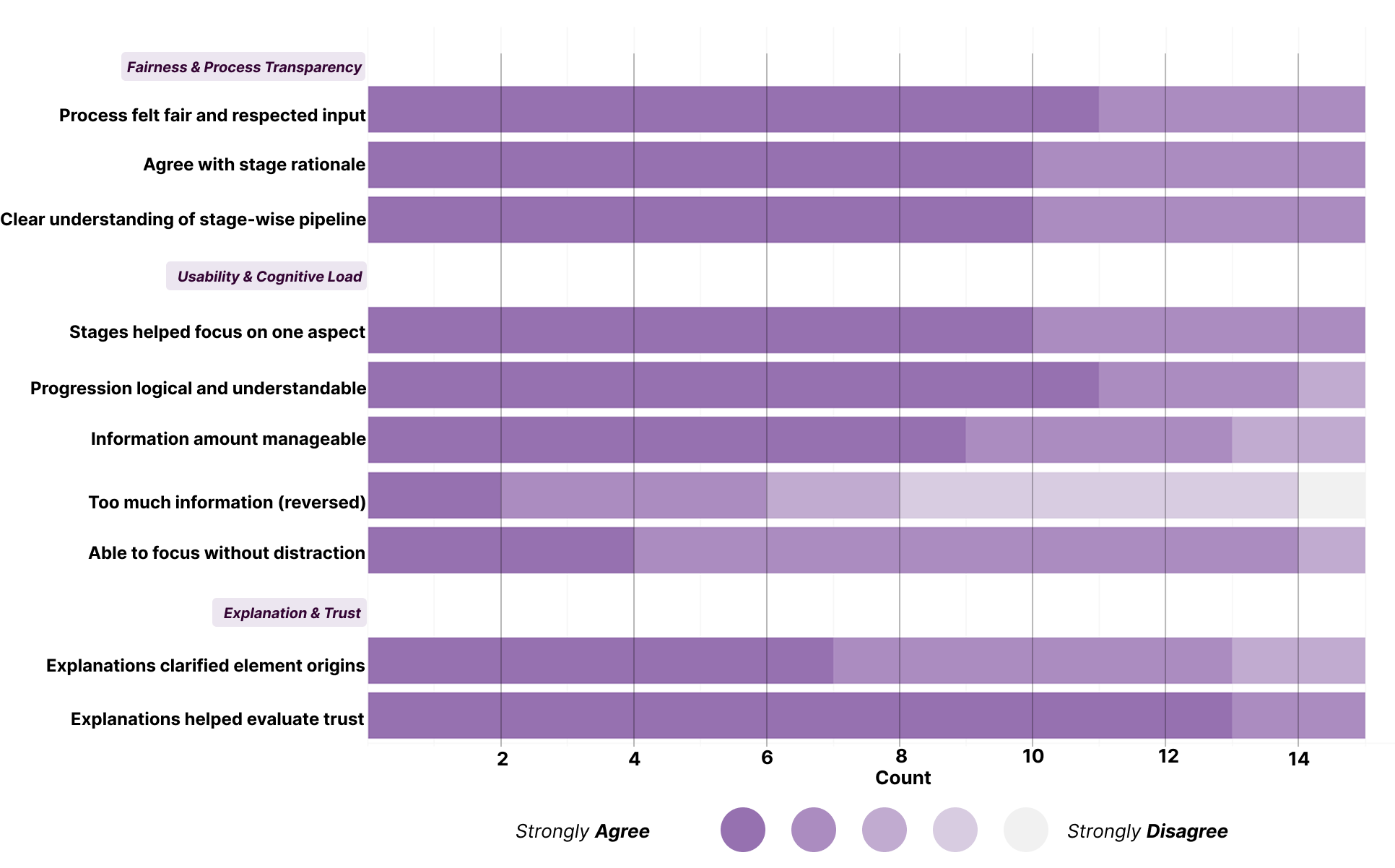}
    \caption{Distribution of participant responses to the 5-point Likert scale questionnaire evaluating the interaction task in LegoUI.}
    \label{fig:inter}
\end{figure*}

This appendix presents the distribution of responses to the 5-point Likert scale questionnaire administered after the interaction task with our LegoUI system, shown in \autoref{fig:inter}.

\newpage
\section{Multi-Tool Port-Generation Task Questionnaire and Response Distribution} \label{A: editing}
\begin{figure*}[h]
    \centering
    \includegraphics[width=1\linewidth]{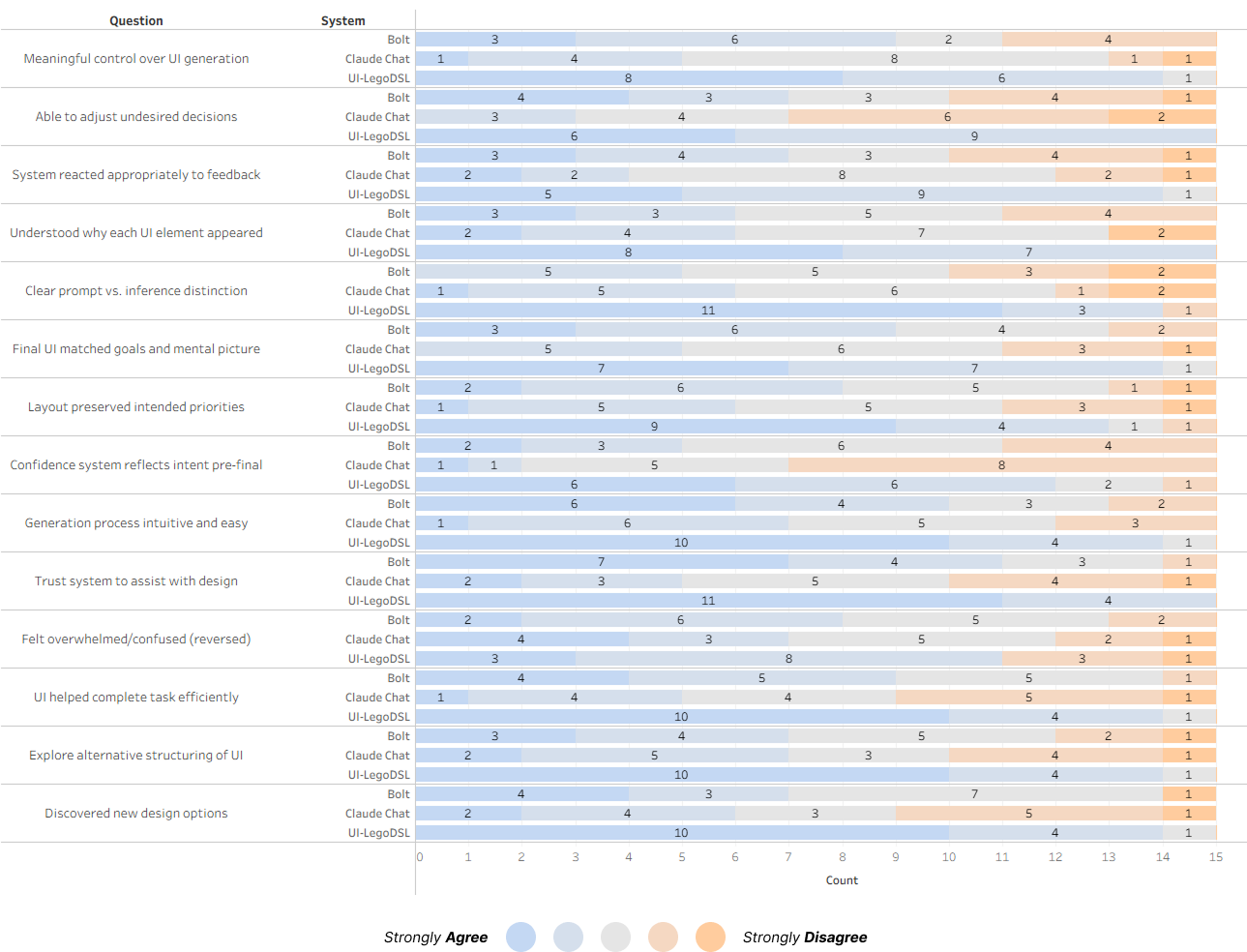}
    \caption{Full distribution of Likert responses for post-generation editing questionnaire items across all five dimensions and three systems.}
    \label{fig:edit}
\end{figure*}

This appendix presents the distribution of responses to the 5-point Likert scale questionnaire administered after the post-generation editing task across all three systems, shown in \autoref{fig:edit}.

\newpage
\section{UI-DSL Grammar and Examples}\label{sec:uidsl}
The UI-DSL organises information into a set of structural specifications that together express the state of the evolving design. Each specification contributes a distinct layer of meaning and records how design decisions accumulate during generation. Individual design decisions recorded within these specifications are represented as DSL items, which serve as the minimal units of reasoning that can be independently addressed and revised during construction. It provides the foundation on which the reasoning stages operate in later sections.

\subsection{Global Level}
The global level records the contextual state that informs how the rest of the specification is constructed and revised.

\noindent\textbf{Intent specification}
The intent specification is a non empty set of intent notes, written as \(\textsc{IntentSpec}=\{\texttt{Note}_1,\ldots\}\) with $(n \ge 1)$. Each note is an atomic intent statement represented as \(\texttt{Note}=\langle id, P_N\rangle\), where \textit{id} uniquely identifies the intent and $P_N$ is a non empty set of provenance entries that articulate the intent statement. All specifications in the UI-DSL carry provenance entries that record how a design decision was introduced or modified. Details of provenance handling are discussed in Section~5.4

\noindent\textbf{Global specification}
The global specification captures contextual settings that apply across the UI-DSL and is written as $\textsc{GlobalSpec}=\texttt{GlobalSetting}\cup\texttt{UserNote}$. Global settings encode stable context and navigation structure, while user notes follow the same note form as in the intent specification and record user commentary or adjustments that may appear at this and subsequent specifications. Navigation is expressed as route mappings of the form $[p_1\rightarrow p_2\mid \phi,\,e]\langle prov\rangle$, denoting a transition from page $p_1$ to page $p_2$ when condition $\phi$ holds and event $e$ occurs. For example, a login transition can be written as $[P_{\text{login}}\rightarrow P_{\text{home}}\mid \text{PageIs}(P_{\text{login}}),\,\text{Submit}(\textit{LoginButton})]\langle prov\rangle$.

\noindent\textbf{Component pool}
The component pool is written as a non empty set of component definitions $\textsc{ComponentPool}=\{C_1,\ldots,C_n\}\cup \mathit{UserNotes}$ with $n\ge 1$. Each component is represented as a tuple $C=(N_c,\tau_c,A_c,P_c)$, denoting its identifier, type, initial attribute assertions, and provenance. Component types range over a fixed set of built-in UI primitives and extensible custom types. Initial attributes are given as a set of attribute assertions $A_c=\{a_1,\ldots,a_k,\beta_1(v_1),\ldots,\beta_m(v_m)\}$, distinguishing state properties from parameterized values. For example, a submit button with an enabled initial state can be written as $C_{\text{submit}}=(N_{\text{submit}},\texttt{Button},\{\texttt{enabled},\texttt{text}(\text{"Submit"})\},P_c)$.

\subsection{Design Dimension Level}
At the design dimension level, the specification records structured design decisions that shape the interface configuration.

\noindent\textbf{Layout specification} 
The layout specification represents page structure as a set of layout regions, written as $\textsc{LayoutSpec}=\{R_1,\ldots,R_n\}\cup \mathit{UserNotes}$ with $n\ge 1$. Each region is written as $R=(N_r,\,L_r,\,C_r,\,P_l)$, denoting its identifier, optional layout description, contained nodes, and provenance. Region content is given as a set of layout nodes $C_r=\{n_1,\ldots,n_k\}$, where each node takes the recursive form $n ::= \texttt{Comp}(c,\,L) \mid \texttt{Group}(g,\,L,\,C_g)$. For example, a vertical grouping of two components can be written as $\texttt{Group}(G_1,\,\{\mathit{dir}=\textit{vertical}\},\,\{\texttt{Comp}(C_1),\,\texttt{Comp}(C_2)\})\langle P_l\rangle$.

\noindent\textbf{Interaction specification}
The interaction specification is written as $\textsc{InteractionSpec}=\{I_1,\ldots,I_n\}\cup \mathit{Patches}\cup \mathit{UserNotes}$. Each interaction rule takes the form $E_c[\Phi]\Rightarrow \mathcal{A}\langle P_i\rangle$, where $E_c$ denotes a trigger event bound to component $c$, $\Phi$ is an optional condition over pages or component attributes, $\mathcal{A}$ is a set of actions over components or pages, and $P_i$ records provenance. For example, a submission-driven navigation can be written as
$$\texttt{Submit}_{\textit{LoginButton}}[\texttt{Has}(\textit{PasswordInput},\textit{filled})]\Rightarrow\{\texttt{Validate}(\textit{LoginForm}),\,\texttt{NavigateTo}(\textit{Dashboard})\}\langle P_i\rangle.$$
When interaction reasoning introduces new components, their placement in the layout is recorded as a patch $P=(c,\,path,\,pos,\,L,\,P_p)$, where $c$ denotes the introduced component, $path$ specifies the target location in the layout hierarchy, $pos$ optionally refines the insertion position, $L$ provides an optional layout description, and $P_p$ records provenance.

\noindent\textbf{Relation specification}
The relation specification is written as a set of relation edges together with user notes, denoted as $\textsc{RelationSpec}=\{R_1,\ldots,R_n\}\cup \texttt{UserNotes}$ with $n\ge 1$. Each relation is written in the form $c_1\rightarrow c_2 : k(a)\langle P_r\rangle$, where the arrow denotes a directed relation from component $c_1$ to component $c_2$, $k(a)$ specifies the relation kind together with the affected attribute, and $P_r$ records provenance. Bidirectional relations are written using a double arrow $c_1\leftrightarrow c_2$. Relation kinds range over a fixed set, including \texttt{functional}, \texttt{state\_dependency}, \texttt{structural}, and \texttt{semantic}. For example, a state dependency between two checkboxes can be written as $$\textit{RememberMeCheckbox}\leftrightarrow \textit{AutoLoginToggle},\texttt{state\_dependency}(\textit{checked})\langle P_r\rangle.$$

\noindent\textbf{Style specification}
The style specification is written as a set of style overrides together with user notes, denoted as $\textsc{StyleSpec}=\{s_1,\ldots,s_n\}\cup \mathit{UserNotes}$. Each override applies a collection of stylistic hints to an existing component or group and is written as $t\mapsto\{h_1,\ldots,h_k\}\langle P_s\rangle$, where $t$ refers to a component or group identifier and $P_s$ records provenance. Each hint $h$ represents a localized stylistic cue used to refine presentation. For example, a visual emphasis applied to a header can be written as $\texttt{Header}\mapsto\{\texttt{palette}(\text{"dark-modern"}),\,\texttt{contrast}(\text{high}),\,\texttt{tag}(\text{"prominent"})\}\langle P_s\rangle$.

\end{document}